\documentclass[12pt]{iopart}
\usepackage{iopams}
\usepackage{bbm}
\usepackage{graphicx}
\usepackage{color}
\usepackage{hyperref}
\hypersetup{
  colorlinks,
  citecolor=blue,
  linkcolor=blue,
  urlcolor=blue}
\usepackage[utf8]{inputenc}
\usepackage[numbers,sort&compress]{natbib}

\newcommand\bbone{\ensuremath{\mathbbm{1}}}
\newcommand{\bra}[1]{\langle #1 \vert}
\newcommand{\bravr}{\langle \boldsymbol{r} \vert}
\newcommand{\braket}[2]{\langle #1 \vert #2 \rangle}
\newcommand{\bravrket}[1]{\langle \boldsymbol{r} \vert #1 \rangle}
\newcommand{\daverage}[1]{\overline{#1}}
\newcommand{\eg}{{\it e.g.\ }}
\newcommand{\Emax}{E_{\mathrm{max}}}
\newcommand{\ie}{{\it i.e.\ }}
\newcommand{\GP}{\mathrm{GP}}
\newcommand{\ket}[1]{\vert #1 \rangle}
\newcommand{\matel}[3]{\langle #1 \vert #2 \vert #3 \rangle}
\newcommand{\phia}{\phi_a}
\newcommand{\phio}{\phi_0}
\newcommand{\qaverage}[1]{\langle #1 \rangle}
\newcommand{\realR}{\mathbb{R}}

\newcommand{\up}{u^\perp}
\newcommand{\upc}{u^{\perp*}}
\newcommand{\vp}{v^\perp}
\newcommand{\vpc}{v^{\perp*}}
\newcommand{\vr}{\boldsymbol{r}}
\newcommand{\vrp}{\boldsymbol{r'}}
\newcommand{\wa}{\tilde{w}}
\newcommand{\wip}{w^\perp}
\newcommand{\wipa}{\tilde{w}^{\perp}}
\begin{document}

\title[Superfluid-insulator transition in weakly interacting disordered Bose
gases]{Superfluid-insulator transition in weakly interacting disordered Bose
gases: a kernel polynomial approach}

\author{J Saliba, P Lugan\footnote{Author to whom any correspondence should be
addressed.} and V Savona}

\address{
Institute of Theoretical Physics, Ecole Polytechnique F\'ed\'erale de Lausanne EPFL,
CH-1015 Lausanne, Switzerland
}

\ead{pierre.lugan@epfl.ch}

\begin{abstract}
An iterative scheme based on the kernel polynomial method is devised for the
efficient computation of the one-body density matrix of weakly interacting Bose
gases within Bogoliubov theory. This scheme is used to analyze the coherence
properties of disordered bosons in one and two dimensions. In the
one-dimensional geometry, we examine the quantum phase transition between
superfluid and Bose glass at weak interactions, and we recover the scaling of
the phase boundary that was characterized using a direct spectral approach by
Fontanesi \etal [Phys.\ Rev.\ A~{\bf 81}, 053603 (2010)]. The kernel polynomial
scheme is also used to study the disorder-induced condensate depletion in the
two-dimensional geometry. Our approach paves the way for an analysis of
coherence properties of Bose gases across the superfluid-insulator transition in
two and three dimensions.
\end{abstract}

% PACS: 05.30.Jp, 03.75.Hh, 05.30.Rt, 64.60.Cn, 02.70.Hm

%\maketitle
\tableofcontents

%%%%%%%%%%%%%%%%%%%%%%%%%%%%%%%%%%%%%%%%%%%%%%%%%%%%%%%%%
\section{Introduction}
%%%%%%%%%%%%%%%%%%%%%%%%%%%%%%%%%%%%%%%%%%%%%%%%%%%%%%%%%

More than 20 years ago, the discovery that superfluidity may be suppressed
in ${}^4\textrm{He}$ adsorbed on porous media~\cite{hertz_marginal_1979,
crooker_superfluidity_1983, reppy_4he_1984, reppy_superfluid_1992} triggered
investigations into the conducting and insulating phases of interacting bosons
in quenched disorder. In this effort to understand what is now known as the
dirty-boson problem~\cite{weichman_dirty_2008}, most studies focused on the
zero-temperature quantum phases, with a variety of approaches including
Luttinger-liquid theory~\cite{giamarchi_localization_1987,
giamarchi_anderson_1988}, general scaling arguments~\cite{ma_strongly_1986,
fisher_boson_1989}, Bogoliubov theory~\cite{lee_bosons_1990,
huang_hard-sphere_1992, nisamaneephong_gaussian_1993, singh_disordered_1994},
strong-coupling expansions~\cite{freericks_strong-coupling_1996}, as well as
numerical calculations with Monte-Carlo~\cite{scalettar_localization_1991,
krauth_superfluid-insulator_1991, prokofev_comment_1998} and DMRG
algorithms~\cite{rapsch_density_1999}. The picture that emerged revealed a rich
interplay of bosonic statistics, disorder, repulsive interactions, and
commensurability effects in the presence of a lattice. The hallmark of this
interplay is the restriction of the superfluid phase to a regime of moderate
interactions and weak disorder, surrounded both at weak and strong interactions
(or strong disorder) by a compressible gapless insulator called Bose
glass~\cite{giamarchi_localization_1987, fisher_boson_1989,
scalettar_localization_1991, krauth_superfluid-insulator_1991}. This picture
holds for bosons both in the continuum and on a lattice, with the difference
that in the case of lattice Bose gases, commensurate fillings give rise to an
additional incompressible Mott-insulator phase at strong coupling, provided the
disorder is bounded~\cite{fisher_boson_1989}. The Bose glass then intervenes
between the superfluid phase and the Mott insulator~\cite{fisher_boson_1989,
rapsch_density_1999, prokofev_comment_1998, pollet_absence_2009,
gurarie_phase_2009}, a feature of the commensurate lattice case that may
nevertheless prove difficult to observe in experiments~\cite{gurarie_phase_2009,
bissbort_stochastic_2010}. At incommensurate fillings or weak interactions, on
the other hand, the lattice case qualitatively resembles the continuous
case~\cite{giamarchi_localization_1987, rapsch_density_1999,
gurarie_phase_2009}. Additionally, in the presence of special symmetries, the
lattice Bose gas may exhibit insulating phases of another kind, such as an
incompressible and gapless Mott glass~\cite{giamarchi_competition_2001,
altman_phase_2004, iyer_mott_2012}. In spite of an important body of available
results, however, the characterization of the generic Bose-glass phase and the
superfluid-insulator transition remains to a large extent an open problem. This
appears to be the case even for the one-dimensional geometry in view of recent
works~\cite{altman_superfluid-insulator_2010, cazalilla_one_2011,
ristivojevic_phase_2012, aleiner_finite-temperature_2010}, which highlight the
challenges related to the connection of the weakly and strongly interacting
regimes and the extension of the ground-state phase diagram to finite
temperatures.

The superfluid-insulator transition of disordered bosons attracted a renewed
interest in the context of ultracold atoms~\cite{damski_atomic_2003}, due to the
high degree of control over disorder, interactions and confining potentials
achieved in these systems~\cite{lewenstein_ultracold_2007, bloch_many-body_2008,
sanchez-palencia_disordered_2010}. While the pioneering experiments with
disordered bosons~\cite{lye_bose-einstein_2005, clement_suppression_2005,
fort_effect_2005, schulte_routes_2005, clement_experimental_2006} aimed at
observing Anderson localization in the noninteracting
limit~\cite{billy_direct_2008, roati_anderson_2008,
jendrzejewski_three-dimensional_2012, piraud_matter_2012}, more recent ones have
provided first results towards a quantitative characterization of the phase
diagram for nonvanishing interactions~\cite{clement_density_2008,
chen_phase_2008, deissler_delocalization_2010, deissler_correlation_2011,
fallani_ultracold_2007, white_strongly_2009, pasienski_disordered_2010}. With
this experimental activity, theoretical investigations also turned to the weakly
interacting regime, which hitherto had been only poorly characterized. The
scaling of the superfluid-insulator phase boundary as a function of the strength
of disorder and interactions was established at the mean-field
level~\cite{lugan_ultracold_2007, nattermann_bose-einstein_2008,
falco_weakly_2009, fontanesi_mean-field_2010, fontanesi_fragmentation_2011}, and
shown to depend in an essential way on the microscopic disorder correlations.
For the 1D geometry, the fragmentation mechanisms driving the transition were
analyzed by means of Bogoliubov theory~\cite{fontanesi_superfluid_2009}, while
universal features of the transition and many-body corrections at intermediate
disorder strengths were worked out with real-space renormalization group (RG)
techniques~\cite{altman_superfluid-insulator_2010,
vosk_superfluid-insulator_2012}. In the latter approach, making contact with
experiments is a challenging task~\cite{vosk_superfluid-insulator_2012}, and the
method itself is not generalized to higher dimensions in a straightforward
way~\cite{iyer_mott_2012}.

To date, the details of the superfluid to Bose-glass transition in dimension
$d>1$ are not well known, and the mechanisms driving the transition are expected
to be more complex than in 1D. The notion of weak links and fragmentation, for
instance, involves connectivity in higher dimensions \ie
percolation~\cite{falco_weakly_2009, pilati_dilute_2010, iyer_mott_2012}.
Moreover, while providing a natural step towards higher dimensions, the 2D case
is particularly interesting in several respect. First, it stands for the
marginal dimension of Anderson localization at the single-particle level, for
the orthogonal and unitary Wigner-Dyson universality
classes~\cite{abrahams_scaling_1979, evers_anderson_2008}. Naively, one would
therefore expect this geometry to be very sensitive to the introduction of
interactions. Second, the clean (disorder-free) weakly interacting system forms
a true condensate at $T=0$, and an algebraic superfluid for
$0<T<T_\mathrm{BKT}$, where $T_\mathrm{BKT}$ is the temperature of the
Berezinskii-Kosterlitz-Thouless
transition~\cite{posazhennikova_colloquium:_2006}. An outstanding question in
this respect is how the zero-temperature superfluid-Bose glass transition
connects to the clean BKT transition as disorder and temperature are varied.
Experiments are ongoing in this regime to characterize the properties of the
disordered Bose gas in 2D~\cite{allard_effect_2012,
beeler_disorder-driven_2012}.

The features of the 2D superfluid to Bose glass transition beyond the mean field
have been addressed recently. In Ref.~\cite{stasinska_glass_2012}, the
Lifshitz-tail physics associated with the deep insulating regime was analyzed by
means of a multi-orbital Hartree-Fock method based on a set of low-lying
single-particle states~\cite{stasinska_glass_2012}. A real-space RG approach was
devised for the 2D dirty-boson problem in Ref.~\cite{iyer_mott_2012}, and
applied to the particle-hole-symmetric case, where the insulating phase is an
incompressible Mott glass instead of a Bose glass. The analysis also emphasized
the possible limitations of the strong-disorder RG in the 2D study. In
Ref.~\cite{astrakharchik_phase_2012} the $T=0$ phase diagram was studied by
means of a weak-disorder expansion of Bogoliubov theory, valid far away from the
transition, and quantum Monte-Carlo calculations. Upon finite-size scaling of
superfluid fractions obtained numerically, the authors concluded in favor of a
smooth crossover between the superfluid and insulating phases. However, the
ad-hoc scaling law used in the analysis may deserve a careful examination, as
the system sizes used in the numerics cover a relatively modest range. Hence,
the development of a method reaching beyond the mean field and affording large
system sizes appears highly desirable in order to fill the existing gaps in the
understanding of the two-dimensional case.

In this contribution, we present a numerical scheme for the efficient
computation of the one-body density matrix of a weakly interacting Bose gas in
the framework of Bogoliubov theory, appropriately extended to account correctly 
for diverging phase fluctuations in low dimensions~\cite{mora_extension_2003}.
The asymptotic behavior of the one-body density matrix determines the superfluid
or insulating behavior of the disordered Bose gas, and thereby allows a
discussion of the phase diagram. Our scheme accomodates arbitrary disorder
strengths, it has no intrinsic limitation in dimensionality, and it admits a
straightforward extension to nonvanishing temperatures within the range of
validity of Bogoliubov theory. The underlying density-phase representation
allows for an accurate description of condensate, quasicondensate and insulating
phases in the limit of large densities for any fixed interaction energy. The key
feature of our approach is that it is based on an iterative scheme called the
kernel polynomial method (KPM)~\cite{weise_kernel_2006}, which allows the
computation of correlation functions in large systems.

The article is organized as follows. Section~\ref{sec:OBDM} provides a reminder
on Bogoliubov theory in the density-phase formulation and on the form of the
one-body density matrix within that framework. The KPM scheme for the
computation of the one-body density matrix is detailed in section~\ref{sec:KPM}.
In section~\ref{sec:results}, we validate our scheme by applying it to the case
of disordered bosons at $T=0$ in 1D and 2D, and by comparing our results with
existing literature. In the 1D geometry, we analyze the destruction of
quasi-long-range order by disorder, and recover the phase diagram obtained
through a direct approach in Refs.~\cite{fontanesi_superfluid_2009,
fontanesi_mean-field_2010}. In the 2D case, we compute the condensate depletion
induced by both interactions and disorder, and compare our findings with those
of Ref.~\cite{muller_condensate_2012}. In section~\ref{sec:conclusion}, we
conclude and discuss extensions of the present work.

%%%%%%%%%%%%%%%%%%%%%%%%%%%%%%%%%%%%%%%%%%%%%%%%%%%%%%%%%
\section{One-body density matrix of weakly interacting Bose gases}
\label{sec:OBDM}
%%%%%%%%%%%%%%%%%%%%%%%%%%%%%%%%%%%%%%%%%%%%%%%%%%%%%%%%%

%%%%%%%%%%%%%%%%%%%%%%%%%%%%%%%%%%%%%%%%%%%%%%%%%%%%%%%%%
\subsection{Long-range and quasi-long-range order}
%%%%%%%%%%%%%%%%%%%%%%%%%%%%%%%%%%%%%%%%%%%%%%%%%%%%%%%%%

We consider a dilute gas of Bose particles described by the many-body
Hamiltonian
%%%%
\begin{equation}
\label{eq:HamiltonianH}
\hat{H}=\int d\vr \left[
\hat{\Psi}^\dagger(\vr) \hat{H_0} \hat{\Psi}(\vr)+\frac{g}{2}
\hat{\Psi}^\dagger(\vr)\hat{\Psi}^\dagger(\vr)\hat{\Psi}(\vr)\hat{\Psi}(\vr)
\right],
\end{equation}
%%%%
where $\hat{\Psi}(\vr)$ is the bosonic field operator, $g>0$ is the coupling
constant parameterizing a repulsive contact interaction, and
$\hat{H}_0=-\frac{\hbar^2}{2m}\nabla_{\vr}^2+V(\vr)$ is the single-particle
Hamiltonian. In the following, the external potential $V$ is a homogeneous
random potential with zero mean, root-mean-square amplitude $\Delta$, and a
correlation length $\eta$ defined as the spatial width of the two-point
correlation function $\daverage{V(\vr)V(\vrp)}$. Here and in the following, the bar
denotes a statistical average over the disorder configurations, while the
brackets $\qaverage{.}$ indicate quantum-mechanical expectation values.

In the weakly interacting regime and close to the ground state, the properties
of the dilute Bose gas are well described by standard Bogoliubov theory in
3D~\cite{pitaevskii_bose-einstein_2003}. In this standard approach, the field
operator $\hat{\Psi}(\vr)$ is split into a classical component $\Psi_0(\vr)$
representing a condensate wave function with well-defined phase and a field
$\delta\hat{\Psi}(\vr)$ describing quantum fluctuations. An effective
Hamiltonian is then derived from the expansion of $\hat{H}$ to second order in
$\delta\hat{\Psi}$. In dimensions $d\leq2$, however, the Mermin-Wagner-Hohenberg
theorem~\cite{mermin_absence_1966, hohenberg_existence_1967} rules out the
presence of a condensate (\ie long-range order~\cite{penrose_bose-einstein_1956,
yang_concept_1962}) at any temperature $T>0$, as well as $T=0$ in 1D. At
sufficienty low temperatures in 2D and $T=0$ in 1D, the Bose gas nevertheless
forms a quasicondensate with a power-law decay of the one-body density matrix
%%%%
\begin{equation}
\label{eq:OBDM}
G(\vr,\vrp)=\qaverage{\hat{\Psi}^\dagger(\vr)\hat{\Psi}(\vrp)}
\end{equation}
%%%%
at large distances ($\Vert\vrp-\vr\Vert\to+\infty$), and exhibits
superfluidity~\cite{cazalilla_one_2011, posazhennikova_colloquium:_2006}. While
the absence of a true condensate in those cases precludes a straightforward
application of standard Bogoliubov theory, the reformulation of the latter in a
density-phase representation~\cite{popov_application_1971,
popov_hydrodynamic_1972} leads to a theory that is free of divergencies, and
which captures the algebraic decay of correlation functions in
quasicondensates~\cite{mora_extension_2003, popov_long-wave_1980}. In this
formulation the field operator writes
$\hat{\Psi}(\vr)=e^{i\hat{\theta}(\vr)}\sqrt{\hat{\rho}(\vr)}$, where
$\hat{\theta}(\vr)$ is a phase operator that is safely defined in the
high-density limit, and $\hat{\rho}(\vr)$ is the density operator. The latter is
split into a classical component $\rho_0(\vr)$ corresponding to the mean-field
condensate (or quasicondensate) density and a fluctuation
$\delta\hat{\rho}(\vr)$. As in standard Bogoliubov theory, an
effective Hamiltonian is derived from the expansion of $\hat{H}$ at leading
order in the fluctuations. At low temperature, such an approach is valid
wherever~\cite{mora_extension_2003}
%%%%
\begin{equation} \label{eq:validity}
\rho_0(\vr)\xi^d\gg1.
\end{equation}
%%%%
In this expression, $d$ is the spatial dimension and $\xi$ is the healing
length, defined here as
%%%%
\begin{equation} \label{eq:healinglength}
\xi=\frac{\hbar}{\sqrt{mU}},
\end{equation}
%%%%
where $U=g\daverage{\rho_0}$ is the average interaction energy. In a disordered
system, the density $\rho_0(\vr)$ may locally assume small values, but the
regime of validity is recovered at identical~$U$ and local interaction energy
$U(\vr)=g\rho_0(\vr)$ for sufficiently large $\daverage{\rho_0}$ (\ie
small~$g$). It is also worth noting that, although the regions of low density
determine the physics of the superfluid-insulator transition, the latter is also
observed within Bogoliubov theory in the limit of asymptotically large average
densities ($\daverage{\rho_0}\to\infty$ and $g\to0$ with constant
$U=g\daverage{\rho_0}$), where the theory is expected to be
exact~\cite{fontanesi_superfluid_2009, fontanesi_quantum_2011}. Many-body
effects beyond Bogoliubov theory arise as subleading terms in an expansion in
terms of the inverse density~\cite{mora_extension_2003,
vosk_superfluid-insulator_2012}. Current experiments with weakly interacting
disordered Bose gases fulfill inequality $\daverage{\rho_0}\xi^d\gg 1$ by a
least one or two orders of magnitude~(see \eg
Refs.~\cite{deissler_delocalization_2010, allard_effect_2012}), so that the
spatial regions where Bogoliubov theory breaks down may be neglected in a good
approximation.

While the relation between superfluidity and condensation or quasicondensation
is rather subtle, the presence of a compressible superfluid appears both
necessary and sufficient for the existence of a condensate or
quasicondensate~\cite{bloch_many-body_2008}. Thus, the long-distance behavior of
the one-body density matrix~(\ref{eq:OBDM}) allows a distinction between
superfluid and insulating phases. In the clean (interacting) 1D system,
$G(r,r')$ decays algebraically at $T=0$. In the clean 2D system at $T=0$, the
one-body density matrix exhibits a plateau at long distances, which
characterizes a true condensate. For $0<T<T_\mathrm{BKT}$, the 2D system is also
an algebraic superfluid with a power-law decay of correlations. In 3D, finally,
the system is a true Bose-Einstein condensate below the critical temperature
$T_\mathrm{BEC}$. All these phases can be distinguished from the normal phases
found at higher temperatures, which exhibit an exponential decay of~$G(\vr,\vrp)$.
Similarly, the complete suppression of superfluidity by disorder at the phase
transition to the Bose glass phase coincides with the destruction of long-range
order or quasi-long-range order, \ie with the emergence of an exponential decay
of the one-body density matrix~\cite{fontanesi_mean-field_2010}. Since the
ground state of the interacting Bose gas is globally extended and $G(\vr,\vrp)$
depends on the disorder configuration, the disordered phases can be
characterized by the long-distance behavior of the statistical average
%%%%
\begin{equation}
\daverage{g_1}(\Vert\vr-\vrp\Vert)=\daverage{g_1(\vr,\vrp)},
\end{equation}
%%%%
where $g_1$ is the \textit{reduced} one-body density matrix
%%%%
\begin{equation}
g_1(\vr,\vrp)=\frac{G(\vr,\vrp)}{\sqrt{\rho(\vr)\rho(\vrp)}},
\end{equation}
%%%%
with $\rho(\vr)$ the total gas density~\cite{mora_extension_2003}.

As anticipated above, $G(\vr,\vrp)$ and $g_1(\vr,\vrp)$ can be calculated in a
density-phase formulation of Bogoliubov theory~\cite{mora_extension_2003}.
Sections~\ref{subsec:Bogoliubov} and~\ref{subsec:OBDM} provide a reminder on
both the number-conserving and nonconserving variations of such a theory. These
results are used in the following sections for the computation of the one-body
matrix with the kernel polynomial method.

%%%%%%%%%%%%%%%%%%%%%%%%%%%%%%%%%%%%%%%%%%%%%%%%%%%%%%%%%
\subsection{Bogoliubov theory in number-conserving and nonconserving approaches}
\label{subsec:Bogoliubov}
%%%%%%%%%%%%%%%%%%%%%%%%%%%%%%%%%%%%%%%%%%%%%%%%%%%%%%%%%

In the ground state, the density $\rho_0$ obeys the Gross-Pitaevskii equation
(GPE)
%%%%
\begin{equation}
\label{eq:GPE}
  [\hat{H}_0+g\rho_0(\vr)]\sqrt{\rho_0(\vr)}=\mu \sqrt{\rho_0(\vr)},
\end{equation}
%%%%
where $\mu$ corresponds to the chemical potential in a grand-canonical
description of the system. The quantum fluctuations and elementary
(quasiparticle) excitations of the Bose gas are described by the field
$\hat{B}(\vr)=\delta\hat{\rho}(\vr)/(2\sqrt{\rho_0(\vr)})+i
\sqrt{\rho_0(\vr)}\hat{\theta}(\vr)$, which obeys the equation of
motion~\cite{mora_extension_2003, petrov_bose-einstein_2003}
%%%%
\begin{equation}
\label{eq:Bmotion}
i \hbar \frac{\partial}{\partial t}
 \left(\begin{array}{c} \hat{B} \\ \hat{B}^\dagger \end{array}\right)
=
\mathcal{L}_{\GP}
 \left(\begin{array}{c} \hat{B} \\ \hat{B}^\dagger \end{array}\right).
\end{equation}
%%%%
Here $\mathcal{L}_{\GP}$ is the standard Bogoliubov operator
%%%%
\begin{equation}
\label{eq:BogoLGP}
\mathcal{L}_{\GP} = \left(\begin{array}{cc}
  \hat{H}_{\GP}+g\rho_0(\vr)-\mu & g\rho_0(\vr) \\
  -g\rho_0(\vr) & -\hat{H}_{\GP}-g\rho_0(\vr)+\mu
\end{array}\right),
\end{equation}
%%%%
with $\hat{H}_{\GP}=\hat{H_0}+g\rho_0(\vr)$. The field $\hat{B}$ admits the
expansion~\cite{mora_extension_2003, castin_low-temperature_1998}
%%%%
\begin{equation}
\label{eq:Bexpansion}
\hat{B}(\vr)=
\sum_j \left[u_j(\vr)\hat{b}_j+v_j^*(\vr)\hat{b}_j^\dagger\right]
-i\sqrt{N_0}\phio(\vr)\hat{Q}_{sb}+\frac{\phia(\vr)}{\sqrt{N_0}}\hat{P}_{sb},
\end{equation}
%%%%
where $\hat{b}_j$ is a bosonic quasiparticle operator, and the two last terms
are explained below. The mode functions $u_j(\vr)=\braket{\vr}{u_j}$ and
$v_j(\vr)=\braket{\vr}{v_j}$ are given by the solutions of the usual
Bogoliubov-de Gennes equations (BdGEs)
%%%%
\begin{equation}
\label{eq:BBdGEs}
\mathcal{L}_{\GP} \left(\begin{array}{c} \ket{u_j} \\ \ket{v_j} \end{array}\right)
=E_j \left(\begin{array}{c} \ket{u_j} \\ \ket{v_j} \end{array}\right)
\qquad(E_j>0).
\end{equation}
%%%%
The operator $\mathcal{L}_{\GP}$ is non-Hermitian, but its eigenvalues are real
in the ground state of Hamiltonian~(\ref{eq:HamiltonianH}). As
${\mathcal{L}_{\GP}}^*=\mathcal{L}_{\GP}$, the components
$u_j(\vr)=\bravrket{u_j}$ and $v_j(\vr)=\bravrket{v_j}$ can be chosen as
real-valued functions. We nevertheless consider the more general case of complex
$u_j$ and $v_j$. For each eigenvector $(\ket{u_j},\ket{v_j})^T$ with eigenvalue
$E_j>0$, the operator $\mathcal{L}_{\GP}$ also has an eigenvector
$(\ket{v_j^*},\ket{u_j^*})^T$ with eigenvalue $-E_j<0$. The adjoint vectors of these
positive and negative eigenvectors are $(\ket{u_j},-\ket{v_j})^T$ and
$(-\ket{v_j^*},\ket{u_j^*})^T$, respectively~\cite{castin_bose-einstein_2001}. The
biorthogonality of both positive and negative solutions of the BdGEs is thus
expressed by the well-known relation
%%%%
\begin{equation}
\label{eq:biorthogonality}
\braket{u_j}{u_{j'}}-\braket{v_j}{v_{j'}}
=\int d\vr [u_j^*(\vr)u_{j'}(\vr)-v_j^*(\vr)v_{j'}(\vr)]
=\delta_{j{j'}}.
\end{equation}
%%%%
The operator $\mathcal{L}_{\GP}$ also has an eigenvector pertaining to the
eigenvalue $E=0$, namely $(\ket{\phio},-\ket{\phio})^T$, where
$\phio(\vr)=\sqrt{\rho_0(\vr)/N_0}$ is the normalized ground state, with
$N_0=\int d\vr \rho_0(\vr)$. As the set of eigenvectors of $\mathcal{L}_{\GP}$
is not complete, an eigenvector $(\ket{\phi_a},\ket{\phi_a})^T$ of
$\mathcal{L}_{\GP}^2$ with eigenvalue zero, such that $\phia(\vr)\in\realR$ and 
$\braket{\phio}{\phia}=1/2$, is added to the set to obtain the closure
relation~\cite{castin_low-temperature_1998}
%%%%
\begin{eqnarray}
\label{eq:Bclosure}
\bbone&=&
\left(\begin{array}{cc}\ket{\phio}\\-\ket{\phio}\end{array}\right)
(\bra{\phia},-\bra{\phia})
+\left(\begin{array}{cc}\ket{\phia}\\\ket{\phia}\end{array}\right)
(\bra{\phio},\bra{\phio})
\nonumber\\
&&+\sum_{j}
\left(\begin{array}{cc}\ket{u_j}\\\ket{v_j}\end{array}\right)
(\bra{u_j},-\bra{v_j})
+\left(\begin{array}{cc}\ket{v_j^*}\\\ket{u_j^*}\end{array}\right)
(-\bra{v_j^*},\bra{u_j^*}).
\end{eqnarray}
%%%%

The $\hat{P}_{sb}$ and $\hat{Q}_{sb}$ terms in Eq.~(\ref{eq:Bexpansion}) account
for fluctuations in the particle number and are responsible for the phase
diffusion of condensates in symmetry-breaking
approaches~\cite{lewenstein_quantum_1996}. These terms do not arise in
number-conserving approaches~\cite{gardiner_particle-number-conserving_1997,
castin_low-temperature_1998, gardiner_number-conserving_2012}, which retain only
fluctuations that are orthogonal to the ground state $\phio(\vr)$. The
field~$\hat{\Lambda}(\vr)$ that describes the orthogonal fluctuations obeys an
equation similar to Eq.~(\ref{eq:Bmotion}):
%%%%
\begin{equation}
\label{eq:Lmotion}
i \hbar \frac{\partial}{\partial t}
 \left(\begin{array}{c} \hat{\Lambda} \\ \hat{\Lambda}^\dagger \end{array}\right)
=
\mathcal{L}
 \left(\begin{array}{c} \hat{\Lambda} \\ \hat{\Lambda}^\dagger \end{array}\right),
\end{equation}
%%%%
where $\mathcal{L}_{\GP}$ has been replaced by~\cite{castin_low-temperature_1998}
%%%%
\begin{equation}
\label{eq:BogoL}
\mathcal{L} = \left(\begin{array}{cc}
  \hat{H}_{\GP}+g\hat{Q}\rho_0(\vr)\hat{Q}-\mu & g\hat{Q}\rho_0(\vr)\hat{Q} \\
  -g\hat{Q}\rho_0(\vr)\hat{Q} & -\hat{H}_{\GP}-g\hat{Q}\rho_0(\vr)\hat{Q}+\mu
\end{array}\right),
\end{equation}
%%%%
and $\hat{Q}=\bbone-\ket{\phio}\bra{\phio}$ projects orthogonally to the ground
state. Equation~(\ref{eq:Bexpansion}) is replaced by the modal expansion
%%%%
\begin{equation}
\label{eq:Lexpansion}
\hat{\Lambda}(\vr)={\sum}_j \up_j(\vr)\hat{b}_j+{\vp_j}^*(\vr)\hat{b}_j^\dagger,
\end{equation}
%%%%
where $\up_j(\vr)$ and $\vp_j(\vr)$ are solution of the modified BdGEs
%%%%
\begin{equation}
\label{eq:LBdGEs}
\mathcal{L} \left(\begin{array}{c} \ket{\up_j} \\ \ket{\vp_j} \end{array}\right)=
E_j \left(\begin{array}{c} \ket{\up_j} \\ \ket{\vp_j} \end{array}\right) \qquad(E_j>0).
\end{equation}
%%%%
The operator $\mathcal{L}$ has the same spectrum as $\mathcal{L}_{\GP}$, and its
positive and negative families of eigenvectors are simply obtained through the
projections $\ket{\up_j}=\hat{Q}\ket{u_j}$ and $\ket{\vp_j}=\hat{Q}\ket{v_j}$,
which leave the biorthogonality relations~(\ref{eq:biorthogonality}) unaffected.
Unlike $\mathcal{L}_{\GP}$, however, the operator $\mathcal{L}$ is
diagonalizable. The zero eigenspace is spanned by the vectors
$(\ket{\phio},0)^T$ and $(0,\ket{\phio})^T$, so that the resolution of identity
writes~\cite{castin_low-temperature_1998}
%%%%
\begin{eqnarray}
\label{eq:Lclosure}
\bbone&=&
\left(\begin{array}{cc}\ket{\phio}\\0\end{array}\right)
(\bra{\phio},0)
+\left(\begin{array}{cc}0\\\ket{\phio}\end{array}\right)
(0,\bra{\phio})
\nonumber\\
&&+\sum_{j}
\left(\begin{array}{cc}\ket{\up_j}\\\ket{\vp_j}\end{array}\right)
(\bra{\up_j},-\bra{\vp_j})
+\left(\begin{array}{cc}\ket{\vpc_j}\\\ket{\upc_j}\end{array}\right)
(-\bra{\vpc_j},\bra{\upc_j}).
\end{eqnarray}
%%%%
As we shall see below, Eqs.~(\ref{eq:GPE}), (\ref{eq:BogoL}) and
(\ref{eq:Lclosure}) [or, equivalently, Eqs.~(\ref{eq:BogoLGP}) and
(\ref{eq:Bclosure})] are all that is needed for the efficient calculation of the
one-body density matrix of disordered Bose gases in the weakly interacting
regime.

%%%%%%%%%%%%%%%%%%%%%%%%%%%%%%%%%%%%%%%%%%%%%%%%%%%%%%%%%
\subsection{One-body density matrix in the density-phase representation}
\label{subsec:OBDM}
%%%%%%%%%%%%%%%%%%%%%%%%%%%%%%%%%%%%%%%%%%%%%%%%%%%%%%%%%

The expression of the one-body density matrix $G(\vr,\vrp)=\langle
\hat{\Psi}^\dagger(\vr)\hat{\Psi}(\vrp)\rangle$ was derived in the density-phase
formalism in Ref.~\cite{mora_extension_2003}. At zero temperature, it can be
cast into the form~\cite{fontanesi_superfluid_2009, fontanesi_quantum_2011}
%%%%
\begin{equation}
\label{eq:G1vperp}
G(\vr,\vrp)=\sqrt{\rho(\vr)\rho(\vrp)}\exp\left[-\frac{1}{2}\sum_j \left|\frac{\vp_j(\vr)}{\sqrt{\rho_0(\vr)}}-\frac{\vp_j(\vrp)}{\sqrt{\rho_0(\vrp)}} \right|^2\right],
\end{equation}
%%%%
where $j$ enumerates the Bogoliubov modes with $E_j>0$. This expression is valid
in the limit of small density and phase fluctuations, which is realized at large
average density $\daverage{\rho_0}$ for any given interaction energy
$U=g\daverage{\rho_0}$. In this limit, one has $\rho(\vr)\simeq\rho_0(\vr)$,
which in the presence of a true condensate amounts to a small condensate
depletion (see section~\ref{subsec:2D}). Note also in this respect
that, owing to the form of the GPE~(\ref{eq:GPE}) and BdGEs~(\ref{eq:LBdGEs}),
the $\vp_j(\vr)$ numerators in the exponent of Eq.~(\ref{eq:G1vperp}) depend
only on the product $U=g\daverage{\rho_0}$ rather than on the coupling constant
$g$ and the average density $\daverage{\rho_0}$ independently. Because of the
denominators, the above exponent thus admits a simple scaling as a function of
density for fixed interaction energy $U$.

Remarkably, expression~(\ref{eq:G1vperp}) accurately describes weakly
interacting Bose gases in any dimension. In particular, it is not plagued by
divergences in low dimensions, and captures the power-law decay of the one-body
density matrix of quasicondensates in 1D Bose gases at
$T=0$~\cite{mora_extension_2003, popov_long-wave_1980, giamarchi_quantum_2004}.
This expression was also used in Ref.~\cite{fontanesi_superfluid_2009} in
conjunction with a numerical diagonalization of the Bogoliubov operator to
analyze the destruction of quasi-long-range order by disorder in the 1D
geometry. While Eq.~(\ref{eq:G1vperp}) involves all Bogoliubov modes, the sum is
typically dominated by the modes of the low-energy phonon regime. However, even
with a restriction to low-energy modes, the calculation of the one-body density
matrix $G(\vr,\vrp)$ through complete or partial diagonalization of the
Bogoliubov operator becomes prohibitive for large system sizes. In the following
section, we present an alternative scheme, based on the kernel polynomial
method~\cite{weise_kernel_2006}, which circumvents the solution of the
Bogoliubov eigenvalue problem and constitutes the main result of the present
work.

%%%%%%%%%%%%%%%%%%%%%%%%%%%%%%%%%%%%%%%%%%%%%%%%%%%%%%%%%
\section{Kernel polynomial scheme for the one-body density matrix}
\label{sec:KPM}
%%%%%%%%%%%%%%%%%%%%%%%%%%%%%%%%%%%%%%%%%%%%%%%%%%%%%%%%%

The kernel polynomial method (KPM)~\cite{silver_densities_1994,
wang_calculating_1994, weise_kernel_2006} is an iterative numerical scheme for
the computation of correlation functions. The KPM bypasses the spectral
decomposition of the operators involved in those correlation functions, which
may be numerically intractable. The KPM technique and some applications have
been recently reviewed in Ref.~\cite{weise_kernel_2006}. In
section~\ref{subsec:KPMbasics} we introduce the elementary aspects of KPM that
are relevant to the present study. In section~\ref{subsec:KPMG1}, we show how a
KPM scheme can be devised to compute the one-body density matrix $G(\vr,\vrp)$ on
the basis of Eq.~(\ref{eq:G1vperp}).

%%%%%%%%%%%%%%%%%%%%%%%%%%%%%%%%%%%%%%%%%%%%%%%%%%%%%%%%%
\subsection{Basics of the kernel polynomial method}
\label{subsec:KPMbasics}
%%%%%%%%%%%%%%%%%%%%%%%%%%%%%%%%%%%%%%%%%%%%%%%%%%%%%%%%%

The core ingredient of KPM is the expansion of correlation functions on
Chebyshev polynomials of the first kind. The latter are orthogonal polynomials
on the interval $I=[-1,1]$, defined by the recurrence relation $ T_{n+1}(x) =
2xT_n(x)-T_{n-1}(x)$ with $T_0(x)=1$ and $T_1(x)=x$. Consider a Hermitian
operator $\hat{X}$ with a discrete or continuous spectrum contained in~$I$, and
the correlation function
%%%%
\begin{equation}
\label{eq:corrfun}
f(\ket{a},\ket{b},x)=\bra{a}\delta(\hat{X}-x)\ket{b},\qquad x\in I.
\end{equation}
%%%%
The latter is a formal writing for
%%%%
\begin{equation}
\label{eq:corrfunspectral}
f(\ket{a},\ket{b},x)=
\sum_{j,\alpha}\delta(x_j-x)
\braket{a}{x_{j,\alpha}}\braket{x_{j,\alpha}}{b},
\end{equation}
%%%%
where $\{\ket{x_{j,\alpha}}\}$ is an orthonormal eigenbasis of $\hat{X}$, which
provides the spectral decomposition $\hat{X}=\sum_{j,\alpha} x_j
\ket{x_{j,\alpha}}\bra{x_{j,\alpha}}$, with $x_j\in I$, and the resolution of
identity
%%%%
\begin{equation}
\label{eq:one}
\bbone=\sum_{j,\alpha} \ket{x_{j,\alpha}}\bra{x_{j,\alpha}}.
\end{equation}
%%%%
The above correlation function has the expansion
%%%%
\begin{equation}
\label{eq:ChebyshevExpansion}
  f(\ket{a},\ket{b},x)=\frac{1}{\pi \sqrt{1-x^2}}\left[ \mu_0(\ket{a},\ket{b})+ 2 \sum_{n=1}^{\infty} \mu_n(\ket{a},\ket{b}) T_n(x) \right],
\end{equation}
%%%%
where the coefficient $\mu_n(\ket{a},\ket{b})$, called Chebyshev moment of order
$n$, is defined as
%%%%
\begin{equation}
\mu_n(\ket{a},\ket{b})=\int_{-1}^1 f(\ket{a},\ket{b},x) T_n(x) dx.
\end{equation}
%%%%
Owing to Eqs.~(\ref{eq:corrfunspectral}) and (\ref{eq:one}), the moments of
$f(\ket{a},\ket{b},x)$ are simply given by
%%%%
\begin{equation}
\mu_n(\ket{a},\ket{b})
=\sum_{j,\alpha} T_n(x_j)\braket{a}{x_{j,\alpha}}\braket{x_{j,\alpha}}{b}
=\bra{a}T_n(\hat{X})\sum_{j,\alpha} \ket{x_{j,\alpha}}\braket{x_{j,\alpha}}{b},
\end{equation}
%%%%
which boils down to the matrix element of a polynomial of $\hat{X}$:
%%%%
\begin{equation}
\mu_n(\ket{a},\ket{b})=\matel{a}{T_n(\hat{X})}{b}.
\end{equation}
%%%%
Instead of calculating $T_n(\hat{X})$ for each new $n$ index, which is
computationally costly, one takes advantage of the recurrence relation between
Chebyshev polynomials and keeps track of two vectors,
$\ket{b_n}=T_n(\hat{X})\ket{b}$ and $\ket{b_{n-1}}=T_{n-1}(\hat{X})\ket{b}$,
with the initialization $\ket{b_0}=\ket{b}$ and $\ket{b_1}=\hat{X}\ket{b}$.
Then, a single application of $\hat{X}$ (matrix-vector multiplication) yields
the new vector $\ket{b_{n+1}}=2\hat{X}\ket{b_n}-\ket{b_{n-1}}$, along with the
next Chebyshev moment $\mu_{n+1}(\ket{a},\ket{b})=\braket{a}{b_{n+1}}.$ In
practice, the expansion~(\ref{eq:ChebyshevExpansion}) is truncated at some
finite order~$N$, and $f(\ket{a},\ket{b},x)$ is approximated by
%%%%
\begin{eqnarray}
%f^{(N)}(\ket{a},\ket{b},x)&=\frac{1}{\pi \sqrt{1-x^2}}\times\nonumber\\
%&\qquad\left[
%g_0^{(N)}\mu_0(\ket{a},\ket{b})
% + 2 \sum_{n=1}^{N-1} g_n^{(N)}\mu_n(\ket{a},\ket{b}) T_n(x)
%\right],
\hspace{-1.5cm}
f^{(N)}(\ket{a},\ket{b},x)&=\frac{1}{\pi \sqrt{1-x^2}}\left[
g_0^{(N)}\mu_0(\ket{a},\ket{b})
 + 2 \sum_{n=1}^{N-1} g_n^{(N)}\mu_n(\ket{a},\ket{b}) T_n(x)
\right],
\end{eqnarray}
%%%%
where the $g_n^{(N)}$ factors are introduced to damp the Gibbs oscillations
caused by the truncation~\cite{weise_kernel_2006}. These factors amount to the
action of a convolution kernel on $f(\ket{a},\ket{b},x)$, whence the name of
KPM. Finally, the functional dependence of $f^{(N)}(\ket{a},\ket{b},x)$ on $x$
is usually efficiently computed for a set of points $x_k\in I$ by using a
discrete cosine transform~\cite{weise_kernel_2006}.

In summary, the KPM offers a simple iterative scheme for the calculation of
correlation functions akin to expression~(\ref{eq:corrfun}), and avoids the
numerical complexity associated with the spectral
representation~(\ref{eq:corrfunspectral}). Let us now examine how this method
can be applied to the Bogoliubov operator for the computation of $G(\vr,\vrp)$.

%%%%%%%%%%%%%%%%%%%%%%%%%%%%%%%%%%%%%%%%%%%%%%%%%%%%%%%%%
\subsection{Calculation of the one-body density matrix}
\label{subsec:KPMG1}
%%%%%%%%%%%%%%%%%%%%%%%%%%%%%%%%%%%%%%%%%%%%%%%%%%%%%%%%%

Expression~(\ref{eq:G1vperp}) for the one-body density matrix involves the
eigenmodes of the Bogoliubov operator $\mathcal{L}$, the spectrum of which lies
on the real line. As all subsequent numerical calculations are carried out with
a finite-difference scheme and finite-size systems, the spectrum of
$\mathcal{L}$ has a compact support $[-\Emax,\Emax]$, where $\Emax$ depends on
the hopping term $t$ associated with the Laplacian in the
finite-difference scheme, the strength of interactions and the disorder
configuration $V(\vr)$. In the calculations presented in
section~\ref{sec:results}, the upper bound $\Emax$ is obtained by solving the
sparse Bogoliubov eigenvalue problem~(\ref{eq:BBdGEs}) for the largest
eigenvalue with a Lanczos method. Taking a slightly larger $\Emax$ to ensure
good KPM convergence at the spectrum boundaries~\cite{weise_kernel_2006}, the
spectrum is mapped to $I$ by the rescaling $\mathcal{L}\to \mathcal{L}/\Emax$.

To exhibit correlation functions akin to Eqs. (\ref{eq:corrfun}) and
(\ref{eq:corrfunspectral}), we cast expression~(\ref{eq:G1vperp}) into the
following form:
%%%%
\begin{equation}
\label{eq:G1F}
G(\vr,\vrp)=\sqrt{\rho(\vr)\rho(\vrp)}\exp\left[
-\frac{1}{2}\int_{0}^1 F(\vr,\vrp,\epsilon) d\epsilon
\right],
\end{equation}
%%%%
with
%%%%
\begin{eqnarray}
\label{eq:bigF}
F(\vr,\vrp,\epsilon)&=
f(\vr,\vr,\epsilon)-f(\vr,\vrp,\epsilon)-f(\vrp,\vr,\epsilon)+f(\vrp,\vrp,\epsilon)
\end{eqnarray}
%%%%
and
%%%%
\begin{equation}
\label{eq:littlef}
f(\vr,\vrp,\epsilon)=
-\sum_k \delta(\epsilon_k-\epsilon)
\frac{\bravrket{\wip_{k,2}}
\braket{\wipa_{k,2}}{\vrp}}{\sqrt{\rho_0(\vr)\rho_0(\vrp)}},
\end{equation}
%%%%
where $\ket{\wip_{k,2}}$ and $\ket{\wipa_{k,2}}$ are the second components of
the eigenvector $(\ket{w_{k,1}},\ket{w_{k,2}})^T$ of~$\mathcal{L}$ and its
adjoint vector $(\ket{\wipa_{k,1}},\ket{\wipa_{k,2}})^T$, respectively. In
Eq.~(\ref{eq:littlef}), the index $k$ runs over the positive ($E_k>0$), null
($E_k=0$) and negative ($E_k<0$) families of eigenvectors, and
$\epsilon_k=E_k/\Emax$. Because of the integration boundaries in
Eq.~(\ref{eq:G1F}), the term $f(\vr,\vrp,\epsilon)$ contributes to the exponent by
%%%%
\begin{equation}
\int_0^1 f(\vr,\vrp,\epsilon) d\epsilon=
-\frac{1}{2 N_0}
+\sum_j\frac{\vp_j(\vr){\vp_j}^*(\vrp)}{\sqrt{\rho_0(\vr)\rho_0(\vrp)}},
\end{equation}
%%%%
where $j$ enumerates the modes with $E_j>0$. The $-1/(2N_0)$ term, which stems
from the zero eigenvector $(0,\ket{\phio})^T$, cancels out of the $f$ sum in Eqs.
(\ref{eq:G1F}) and (\ref{eq:bigF}), so that Eq.~(\ref{eq:G1F}) is indeed
equivalent to Eq.~(\ref{eq:G1vperp}).

The modes with $E_k\leq 0$ are included in sum~(\ref{eq:littlef}) to use the
resolution of identity [see Eq.~(\ref{eq:Lclosure})]
%%%%
\begin{equation}
\bbone=\sum_k
(\ket{\wip_{k,1}},\ket{\wip_{k,2}})^T
(\bra{\wipa_{k,1}},\bra{\wipa_{k,2}}).
\end{equation}
%%%%
Indeed, rewriting Eq.~(\ref{eq:littlef}) as
%%%%
\begin{eqnarray}
f(\vr,\vrp,\epsilon)&=-\sum_k \delta(\epsilon_k-\epsilon) \frac{
(0,\bravr)
(\ket{\wip_{k,1}},\ket{\wip_{k,2}})^T
(\bra{\wipa_{k,1}},\bra{\wipa_{k,2}})
(0,\ket{\vrp})^T
}{\sqrt{\rho_0(\vr)\rho_0(\vrp)}},\nonumber\\
\label{eq:littleflong}
\end{eqnarray}
%%%%
we find that the Chebyshev moments of $f(\vr,\vrp,\epsilon)$ are
%%%%
\begin{equation}
\label{eq:munTn}
\mu_n(\vr,\vrp)=-
\frac{(0,\bravr)}{\sqrt{\rho_0(\vr)}}
T_n\left(\frac{\mathcal{L}}{\Emax}\right)
\frac{(0,\ket{\vrp})^T}{\sqrt{\rho_0(\vrp)}},
\end{equation}
%%%%
and those of $F(\vr,\vrp,\epsilon)$ follow as
%%%%
\begin{equation}
\label{eq:Mn}
M_n(\vr,\vrp)=\mu_n(\vr,\vr)-\mu_n(\vr,\vrp)-\mu_n(\vrp,\vr)+\mu_n(\vrp,\vrp).
\end{equation}
%%%%
Finally, there is no need for a Chebyshev inversion by discrete cosine
transform, as the expansion~(\ref{eq:ChebyshevExpansion}) can be integrated
analytically on $[0,1]$, and we obtain
%%%%
\begin{equation}
\int_0^1 F(\vr,\vrp,\epsilon) d\epsilon=
\frac{M_0(\vr,\vrp)}{2}
+\frac{2}{\pi}\sum_{p=0}^{\infty}
\frac{(-1)^p M_{2p+1}(\vr,\vrp)}{2p+1}.
\label{eq:Final_Exponent}
\end{equation}
%%%%
Note that the contributions of all even moments except $M_0(\vr,\vrp)$ are
integrated out on $[0,1]$. The moments $\mu_n(\vr,\vrp)$ are calculated iteratively
following the recurrence scheme outlined in section~\ref{subsec:KPMbasics},
for the four $(\vr,\vrp)$ pairs in $M_n(\vr,\vrp)$. This requires only two Chebyshev
sequences as, for instance, $T_n(\mathcal{L}/\Emax)(0,\ket{\vrp})^T$ may be used
to compute $\mu_n(\vr,\vrp)$ and $\mu_n(\vrp,\vrp)$. When the Chebyshev iterations are
truncated at order $N$, the reduced one-body density matrix is approximated by
%%%%
\begin{equation}
\label{eq:reducedGN}
\frac{G^{(N)}(\vr,\vrp)}{\sqrt{\rho(\vr)\rho(\vrp)}}=
\exp\left[
-\frac{g_0^{(N)}}{4}M_0(\vr,\vrp)
-\sum_{p=0}^{\lfloor \frac{N}{2}-1\rfloor }
\frac{(-1)^p g_{2p+1}^{(N)}}{(2p+1)\pi}M_{2p+1}(\vr,\vrp)
\right],
\end{equation}
%%%%
where the $g_n^{(N)}$ are convolution kernel factors (see
section~\ref{subsec:KPMbasics}). We found the standard Jackson
kernel~\cite{weise_kernel_2006} to be suitable in this scheme.

The Chebyshev iterations based on Eq.~(\ref{eq:munTn}) involve projections
orthogonally to the ground state by means of
$\hat{Q}=\bbone-\ket{\phio}\bra{\phio}$. Interestingly, the Bogoliubov operator
$\mathcal{L}$ may be replaced by $\mathcal{L}_{\GP}$ in the iterations, so that
projections are not necessary. This provides a further simplification of the KPM
calculation of the one-body density matrix. Indeed, upon expansion with
$v_j^\perp(\vr)=v_j(\vr)-\braket{\phio}{v_j}\phio(\vr)$, one easily sees that
Eq.~(\ref{eq:G1vperp}) also writes as
%%%%
\begin{equation}
\label{eq:G1v}
G(\vr,\vrp)=\sqrt{\rho(\vr)\rho(\vrp)}\exp\left[-\frac{1}{2}\sum_j \left|\frac{v_j(\vr)}{\sqrt{\rho_0(\vr)}}-\frac{v_j(\vrp)}{\sqrt{\rho_0(\vrp)}} \right|^2\right].
\end{equation}
%%%%
The function $F_{\GP}(\vr,\vrp,\epsilon)$ is defined as the analogue of
$F(\vr,\vrp,\epsilon)$ in Eqs.~(\ref{eq:G1F}) and (\ref{eq:bigF}), with four
terms of the form
%%%%
\begin{equation}
\label{eq:littlefGP}
f_{\GP}(\vr,\vrp,\epsilon)=-\sum_{k'}
\delta(\epsilon_{k'}-\epsilon)
\frac{\bravrket{w_{k',2}}
\braket{\wa_{k',2}}{\vrp}}{\sqrt{\rho_0(\vr)\rho_0(\vrp)}},
\end{equation}
%%%%
where $k'$ enumerates the eigenmodes of $\mathcal{L}_{\GP}$ with $E_{k'}\neq0$.
Owing to the closure relation~(\ref{eq:Bclosure}), the Chebyshev moments of
$f_{\GP}(\vr,\vrp,\epsilon)$ read
%%%%
\begin{equation}
\label{eq:munTnGP}
\mu_n^{\GP}(\vr,\vrp)=
-\frac{(0,\bravr)}{\sqrt{\rho_0(\vr)}}
T_n\left(\frac{\mathcal{L}_{\GP}}{\Emax}\right)
\frac{(0,\ket{\vrp})^T}{\sqrt{\rho_0(\vrp)}}
+R_n^{(1)}(\vr,\vrp)+R_n^{(2)}(\vr,\vrp),
\end{equation}
%%%%
where
%%%%
\begin{eqnarray}
\label{eq:Rn1}
R_n^{(1)}(\vr,\vrp)=
-\frac{\phia(\vrp)}{\sqrt{\rho_0(\vr)\rho_0(\vrp)}}
(0,\bravr)\,
T_n\left(\frac{\mathcal{L}_{\GP}}{\Emax}\right)
\left(\begin{array}{cc}\ket{\phio}\\-\ket{\phio}\end{array}\right)\\
\label{eq:Rn2}
R_n^{(2)}(\vr,\vrp)=
+\frac{\phio(\vrp)}{\sqrt{\rho_0(\vr)\rho_0(\vrp)}}
(0,\bravr)\,
T_n\left(\frac{\mathcal{L}_{\GP}}{\Emax}\right)
\left(\begin{array}{cc}\ket{\phia}\\\ket{\phia}\end{array}\right).
\end{eqnarray}
%%%%
Given that $\mathcal{L}_{\GP}(\ket{\phio},-\ket{\phio})^T=(0,0)^T$ and
$\mathcal{L}_{\GP}(\ket{\phia},\ket{\phia})^T=\alpha(\ket{\phio},-\ket{\phio})^T$
with constant~$\alpha$, as detailed in Ref.~\cite{castin_low-temperature_1998},
we find
%%%%
\begin{eqnarray}
R_{2p}^{(1)}(\vr,\vrp)&=(-1)^{p}\frac{\phia(\vrp)}{\sqrt{N_0 \rho_0(\vrp)}} \label{eq:R21}\\
R_{2p+1}^{(1)}(\vr,\vrp)&=0\\
R_{2p}^{(2)}(\vr,\vrp)&=(-1)^{p}\frac{\phia(\vr)}{\sqrt{N_0 \rho_0(\vr)}} \\
R_{2p+1}^{(2)}(\vr,\vrp)&=(-1)^{p+1}\frac{(2p+1)\alpha}{N_0 \Emax}.\label{eq:R212}
\end{eqnarray}
%%%%
These $R_n^{(i)}(\vr,\vrp)$ terms cancel out of the sum
%%%%
\begin{equation}
M_n^{\GP}(\vr,\vrp)=
\mu_n^{\GP}(\vr,\vr)-\mu_n^{\GP}(\vr,\vrp)-\mu_n^{\GP}(\vrp,\vr)+\mu_n^{\GP}(\vrp,\vrp).
\end{equation}
%%%%
Hence, the comparison of Eqs.~(\ref{eq:munTn}) and (\ref{eq:munTnGP}) shows that
$\mathcal{L}_{\GP}$ can be used instead of $\mathcal{L}$ in the Chebyshev
iterations underlying Eq.~(\ref{eq:reducedGN}).

The expressions (\ref{eq:munTn}), (\ref{eq:Mn}) and (\ref{eq:reducedGN}) are the
main results of this section. They provide an iterative scheme for the
calculation of the reduced one-body density matrix of a weakly interacting Bose
gas, once the solution of the GPE~(\ref{eq:GPE}) is given. In the following
section, this scheme is applied in various geometries, with and without
disorder.

Remarkably, the above approach can be extended in a straightforward way to
nonzero temperatures within the framework of Bogoliubov theory. In addition to
the $v_j(\vr)$ terms, the thermal $G(\vr,\vrp)$ also contains contributions from
the Bogoliubov components $u_j(\vr)$, which can be calculated through additional
KPM iterations. As the $u_j$ and $v_j$ terms are weighted by Bose-Einstein
occupation factors, the integration in Eq.~(\ref{eq:Final_Exponent}) no longer
has a simple analytical solution. However, this analytical step may be replaced
by a discrete cosine transform at low computational expense.

Our results also show that the operator involved in the correlation function and
the KPM iterations need not be Hermitian. Some illustrations of this fact can be
found in the literature, with special cases such as the computation of retarded
Green's functions~\cite{weise_kernel_2006} or the solution of
\textit{generalized} Hermitian eigenvalue problems~\cite{roder_kernel_1997}. The
Bogoliubov operator $\mathcal{L}$ and $\mathcal{L}_{\GP}$ provide two other
interesting examples in that context. First, $\mathcal{L}$ is diagonalizable,
albeit non-Hermitian, and its eigenvalues are real. As a consequence, the
eigenvectors and their adjoints form a complete biorthogonal set that can be
used for the closure relation, and there is no need for a twofold KPM iteration
to retrieve the spectral information on a compact set of the complex plane.
Second, $\mathcal{L}_{\GP}$ is not diagonalizable, and yet its
\textit{generalized} eigenvectors (and their adjoints) can be used for the
closure relation. In the derivation of Eqs.~(\ref{eq:R21}) to~(\ref{eq:R212}),
we took advantage of the fact that $(\ket{\phio},-\ket{\phio})^T$ and
$(\ket{\phia},\ket{\phia})^T$ are generalized eigenvectors of low order, so that
the action of $T_n(\mathcal{L}_{\GP})$ can be evaluated easily.

%%%%%%%%%%%%%%%%%%%%%%%%%%%%%%%%%%%%%%%%%%%%%%%%%%%%%%%%%
\section{Application to disordered bosons}
\label{sec:results}
%%%%%%%%%%%%%%%%%%%%%%%%%%%%%%%%%%%%%%%%%%%%%%%%%%%%%%%%%

We now employ the KPM scheme introduced above to analyze the asymptotic behavior
of the one-body density matrix of disordered bosons in 1D and 2D. While the
approach of the previous sections is general, we consider here a Gaussian random
potential $V$ with Gaussian correlation function
%%%%
\begin{equation} \label{eq:gaussian}
\daverage{V(\vr)V(\vrp)}=\Delta^2 e^{-(\vr-\vrp)^2/(2\eta^2)}.
\end{equation}
%%%%
The numerical procedure for the computation of $g_1(\vr,\vrp)$ is the following. The
continuum problem~(\ref{eq:HamiltonianH}) is represented on a finite volume
$L^d$ in a finite-difference scheme with lattice spacing $\ell=L/n_\ell$. To
emulate the continuum limit, the hopping term $t=\hbar^2/(2m\ell^2)$ is chosen
to be much larger than all the other energy scales of the problem. In all the
calculations presented here, the correlation length of the disorder is taken to
be $\eta=4\ell$, which is sufficient for our purposes. The correlation length
$\eta$ and the associated energy
%%%%
\begin{equation} \label{eq:Ec}
E_c=\frac{\hbar^2}{2m\eta^2}
\end{equation}
%%%%
are used as reference scales throughout this section, even in the absence of
disorder ($\Delta=0$). For each configuration $V(\vr)$, the ground-state
solution $\sqrt{\rho_0(\vr)}$ of the GPE~(\ref{eq:GPE}) and the corresponding
chemical potential $\mu$ are computed through imaginary-time propagation with a
standard Crank-Nicolson scheme, at fixed particle number $N_0=\int d\vr
\rho_0(\vr)$. We denote by
%%%%
\begin{equation} \label{eq:U}
U=g \frac{N_0}{L^d}
\end{equation}
%%%%
the average mean-field interaction energy. Periodic boundary conditions are
imposed on the GPE and on the BdGEs~(\ref{eq:BBdGEs}). The Bogoliubov operator
$\mathcal{L}_{\GP}$ is set up on the basis of $V(\vr)$ and $\rho_0(\vr)$. Then,
$g_1(\vr,\vrp)$ may be calculated with the KPM iteration detailed in the last
section for any $(\vr,\vrp)$ pair.

%%%%%%%%%%%%%%%%%%%%%%%%%%%%%%%%%%%%%%%%%%%%%%%%%%%%%%%%%
\subsection{Superfluid to Bose glass transition in one dimension}
\label{subsec:1D}
%%%%%%%%%%%%%%%%%%%%%%%%%%%%%%%%%%%%%%%%%%%%%%%%%%%%%%%%%

In Refs.~\cite{fontanesi_superfluid_2009, fontanesi_mean-field_2010}, the
destruction of quasi-long-range order was used as a signature of the superfluid
to Bose glass transition at $T=0$ in 1D, and this criterion was used to draw the
quantum phase diagram on the basis of the asymptotic behavior of the (averaged)
reduced one-body density matrix $\daverage{g_1}(|r-r'|)$. Here, we use the 1D
setting as a testbed for the KPM approach described above.

\begin{figure}
\centering
\includegraphics[width=0.55\linewidth,angle=-90]{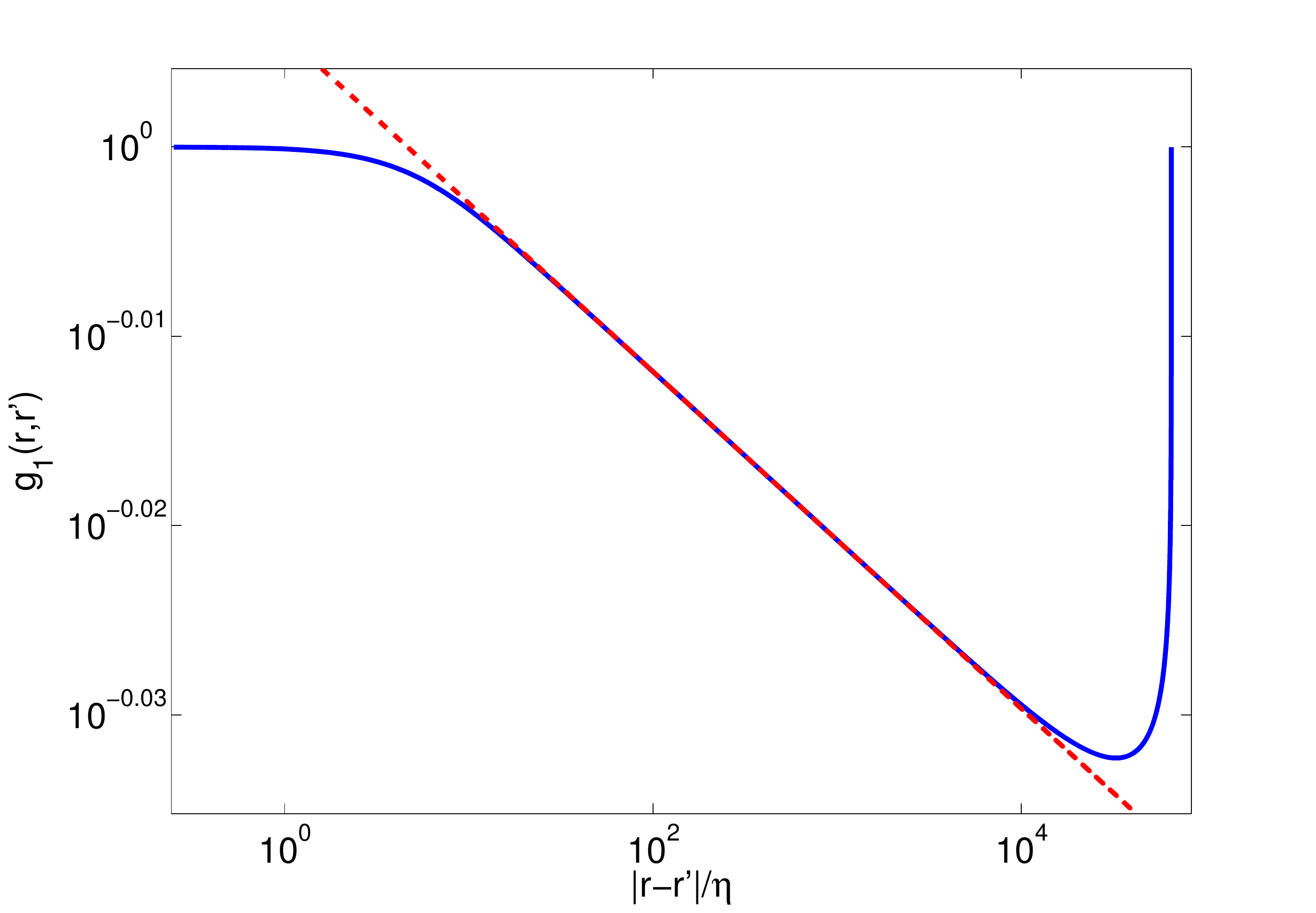}
\caption{\label{fig:1D_clean_g1}
Reduced one-body density matrix $g_1(r,r')$ in the absence of disorder for a 1D
system of length $L=2^{16}\, \eta$, with interaction strength $U=0.10\, E_c$
(\ie $\xi\simeq4.5\,\eta$). The blue solid line is the result of the KPM
calculation. The red dashed line is the asymptotic power law given by
Eq.~(\ref{eq:1Dcleang1}) with $\rho\simeq\rho_0$.
}
\end{figure}

In the absence of disorder, $g_1$ is expected to decay at large distances with a
power law given by~\cite{popov_theory_1972,
mora_extension_2003}:
%%%%
\begin{equation} \label{eq:1Dcleang1}
g_1(r,r')\simeq\left(\frac{e^{2-C}\xi}{4|r-r'|}\right)^\frac{1}{2\pi\rho\xi},
\qquad |r-r'|\gg \xi,
\end{equation}
%%%%
where $C=0.57721\dots$ is Euler's constant, the density $\rho$ can be
approximated by $\rho_0$ within our Bogoliubov approach, and $\xi$ is the
healing length defined in Eq.~(\ref{eq:healinglength}). Figure~\ref{fig:1D_clean_g1} 
%%%%
%%%%
shows the result of a KPM calculation for a system of length $L=2^{16} \eta$,
and the excellent agreement obtained with the power law~(\ref{eq:1Dcleang1}) for
$|r-r'|\gtrsim\xi$, \ie in its regime of validity. The regrowth observed at
large $|r-r'|$ is due to the periodic boundary conditions, and does not affect
significantly the data for $|r-r'|\lesssim L/4$. In all the subsequent analyses,
we retain only this range for determining the asymptotic behavior of $g_1$. Note
the large system size achieved in this computation. In this homogeneous case, a
single KPM iteration suffices for the computation of $g_1(r,r')$. The number of
moments~$N$ required to resolve all individual Bogoliubov modes in the phonon
regime and achieve a good convergence of the KPM result grows linearly with the
system size $n_\ell=L/\ell$. The required storage space is also of a few
$n_\ell$. This has to be compared to the storage space and computation time
required for a full diagonalization of $\mathcal{L}_{\GP}$, which scale as
$n_\ell^2$ and $n_\ell^3$ respectively~\cite{weise_kernel_2006}.

\begin{figure}
\centering
\includegraphics[width=0.55\linewidth,angle=-90]{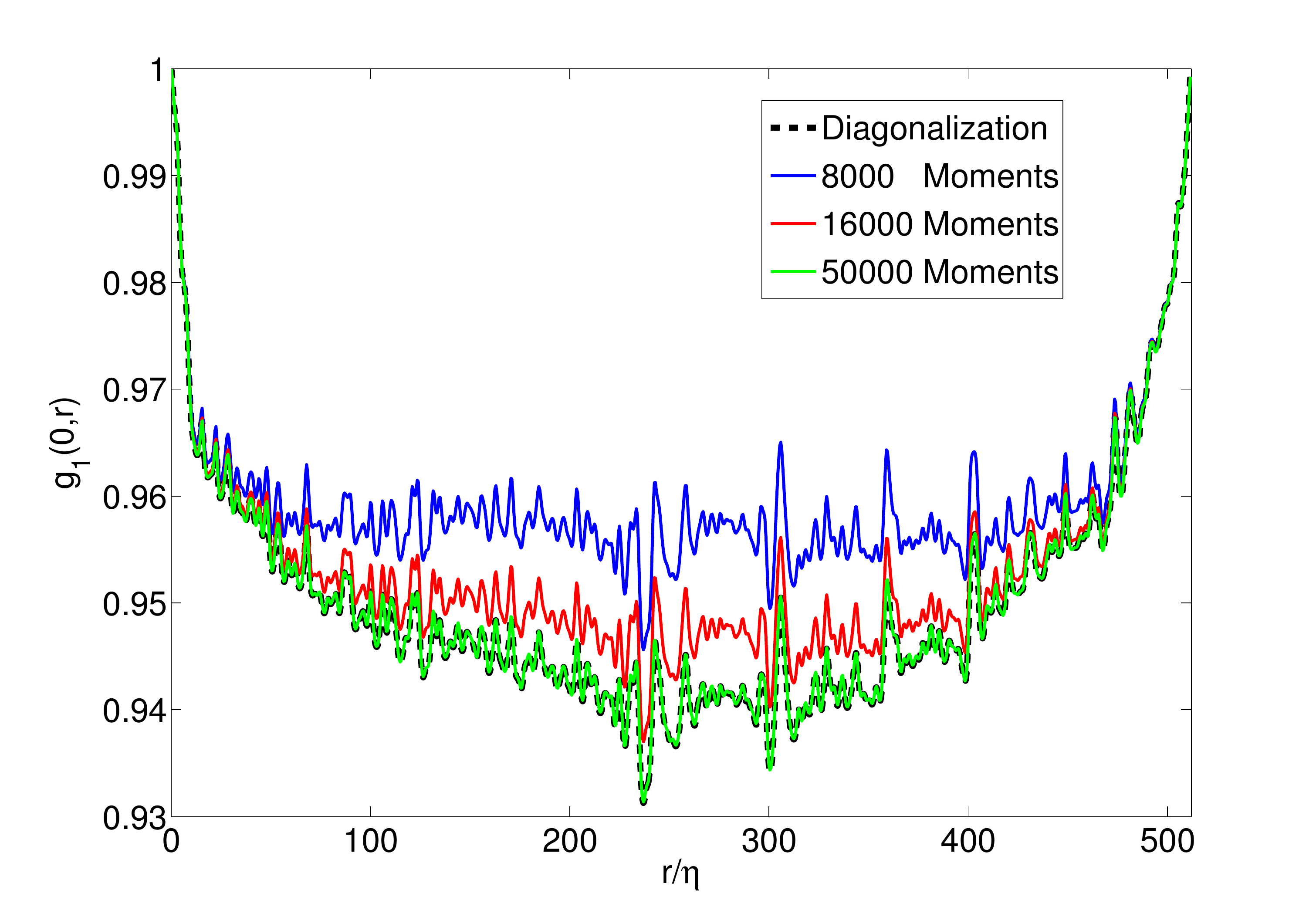}
\caption{\label{fig:1D_disorder_oneconfig}
Reduced one-body density matrix $g_1(0,r)$ for a single disorder configuration,
in a 1D system of length $L=512\,\eta$. The interaction strength is $U=g
N_0/L=1.12\, E_c$, and the disorder strength is $\Delta=0.8\, E_c$. The black
dashed line shows the result obtained from the complete diagonalization of the
Bogoliubov operator $\mathcal{L}_{\GP}$. The other curves are the KPM results
obtained for various numbers of moments. The curve for 50000 moments
(green solid line) is undistinguishable on this scale from the result obtained
from complete diagonalization.
}
\end{figure}

For $\Delta>0$, the reduced one-body density matrix $g_1(r,r')$ depends on the
disorder configuration, and is no longer translation invariant.
Fig.~\ref{fig:1D_disorder_oneconfig} shows the behavior of $g_1(0,r)$ for a
single disorder configuration, and the results obtained with the KPM scheme for
various moment numbers $N$ [see Eq.~(\ref{eq:reducedGN})]. When $N$ is
sufficiently high, the KPM results coincide with those obtained from complete
diagonalization, and faithfully reproduce the spatial details of $g_1(r,r')$. As
a general trend, we also observe that the KPM estimates of $g_1(r,r')$ converge
from above with increasing $N$. This can be attributed to the fact that low $N$
values imply poor spectral resolution, and hence are not able to resolve the
small energy scales associated with the long-distance decay of $g_1$, such as
for example the vanishing energy separation that can arise when Bogoliubov modes
are strongly localized in different spatial regions. While the number of moments
required for convergence scales linearly with the system size in the clean case,
we also observed that this scaling is slightly faster than linear in the
disordered case.

%%%%
%%%%

\begin{figure}[!t]
%\centering
\flushright
\includegraphics[width=0.355\linewidth, trim=0cm 0cm 0cm 2cm,
clip=true, angle=-90]{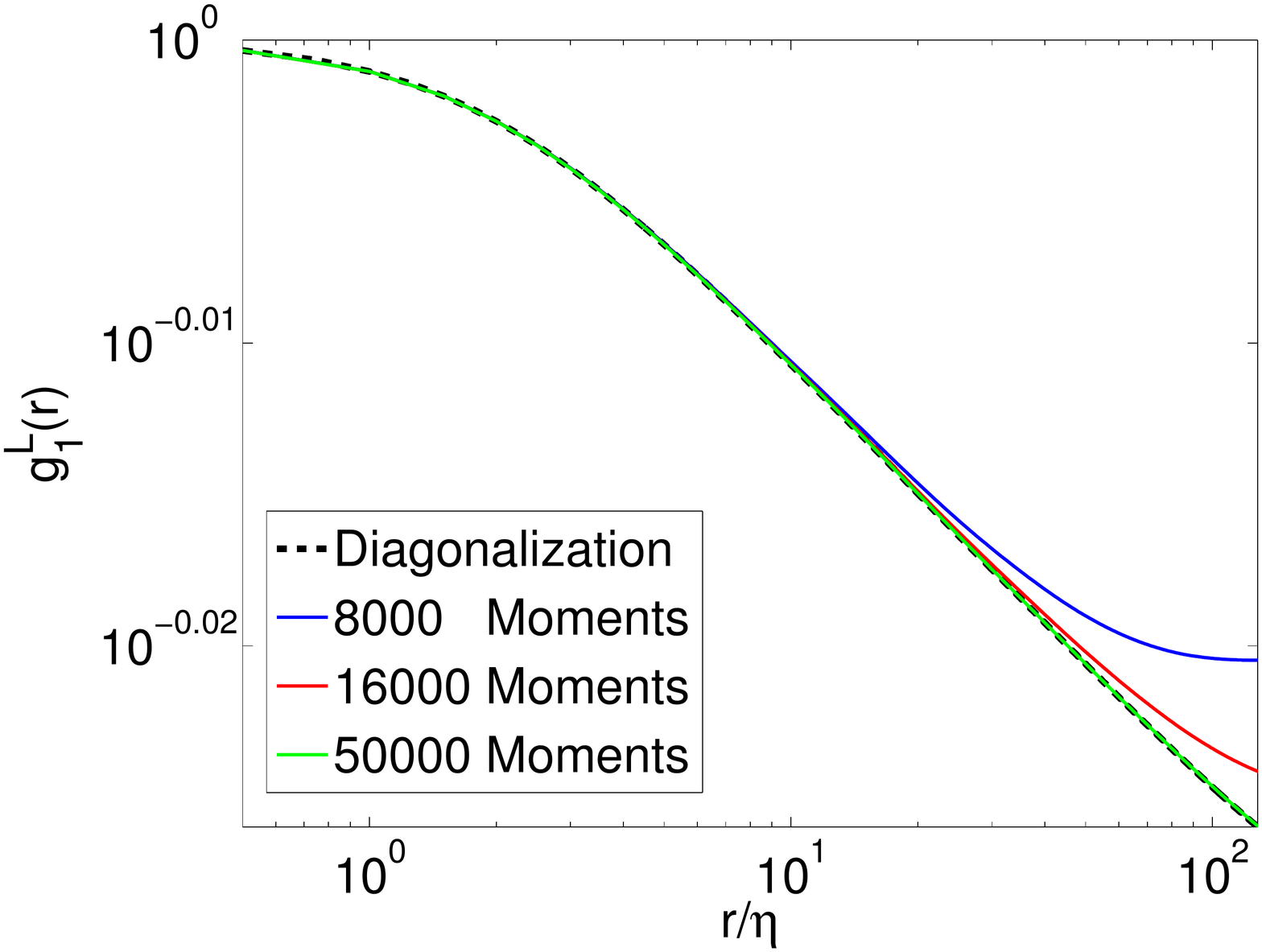}
\includegraphics[width=0.355\linewidth, trim=0cm 0cm 0cm 2cm, 
clip=true, angle=-90]{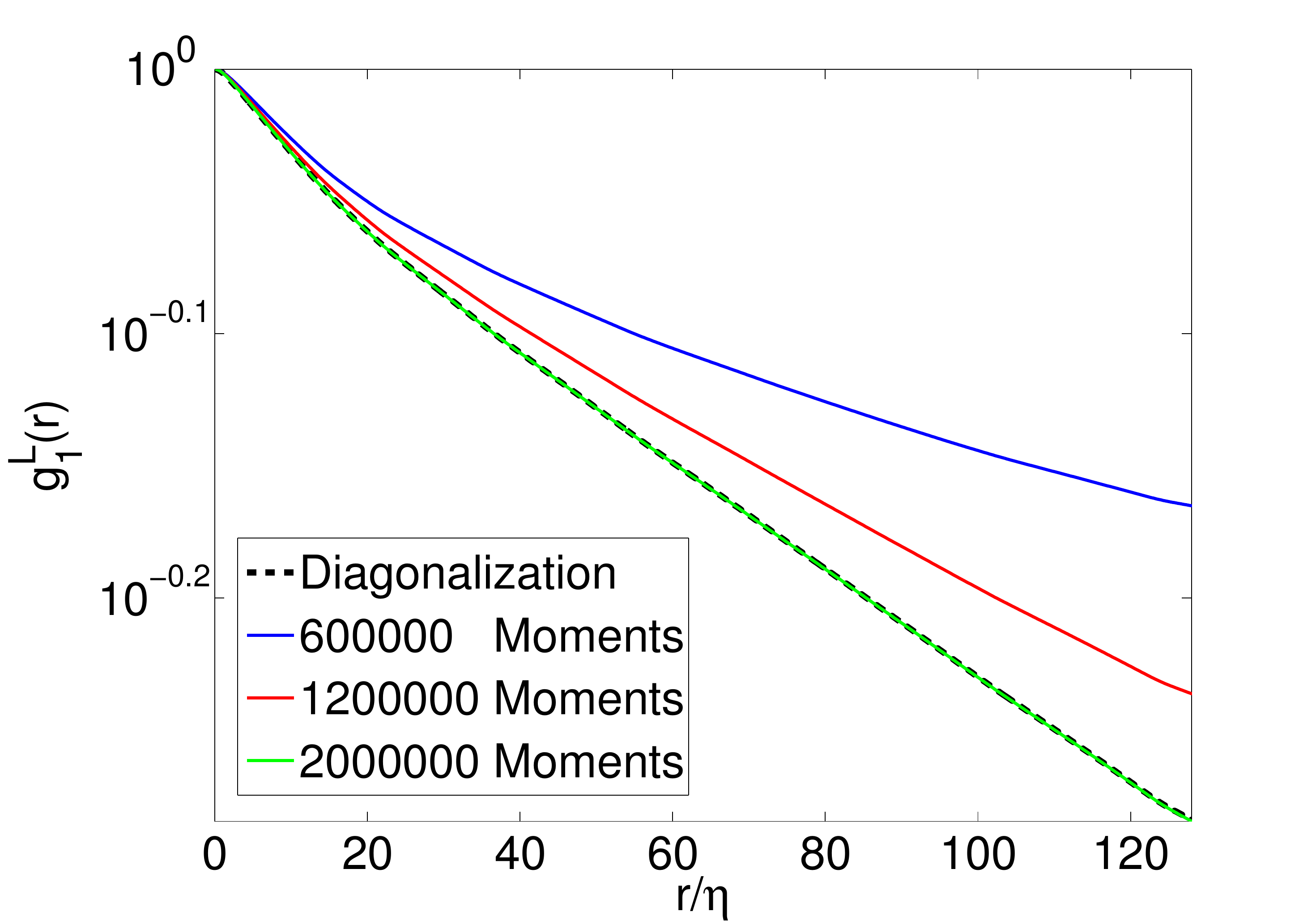}
\caption{ \label{fig:BG_SF}
Spatial average $g_1^L(r)$ for a single disorder configuration with amplitude
$\Delta=0.8\, E_c$ and two interaction energies $U$, in a 1D system of length
$L=512\,\eta$. The case $U=1.12\, E_c$ (left panel) exhibits a power-law decay
of $g_1^L$, while the case $U=0.3\,E_c$ (right panel) shows an exponential
decay for $r\gtrsim20\,\eta$. The left panel corresponds
to the spatial average obtained with the parameters and the disorder
configuration of Fig.~\ref{fig:1D_disorder_oneconfig}.
}
\end{figure}

After averaging over disorder, the one-body density matrix exhibits either a
power-law or an exponential decay a large distances, depending on the strength
of interactions and disorder. For long-enough systems a similar behavior may
already be observed qualitatively with a single disorder configuration in the
spatial average
%%%%
\begin{equation}\label{eq:g1L}
g_1^L(r)=\frac{1}{L}\int_0^L g_1(r',r'+r) dr'.
\end{equation}
%%%%
Fig.~\ref{fig:BG_SF} shows spatial averages computed with the KPM scheme for two
sets of parameters in a system of length $L=512\,\eta$. We display here the
range $r\lesssim L/4$, and the crossover to the long-distance behavior is
visible on both panels. The linear behavior found for $U=1.12\, E_c$ on the
double logarithmic scale indicates a power-law decay of the averaged $g_1$,
corresponding to the superfluid phase. For $U=0.3\, E_c$, on the other hand, we
find an exponential decay indicating a Bose glass. For the same system size, the
convergence of the KPM result in the insulating phase requires a higher number
of moments than in the superfluid phase. This may be attributed to the increase
of the Bogoliubov density of states near zero energy and to a reduced level
repulsion. Indeed, as the disorder increases, the system turns progressively
into a collection of superfluid puddles separated by high potential barriers.
As a consequence, the efficiency of the KPM scheme is reduced deep
in the insulating regime, where the number of moments required for convergence
typically becomes very large. In the superfluid regime and in the parameter
range of interest around the superfluid-insulator transition, however, our KPM
approach converges quickly and outperforms complete diagonalization in terms of
memory usage and computation time already for system sizes as moderate as those
used in Fig.~\ref{fig:BG_SF}.

%%%%
%%%%

\begin{figure}
\centering
\includegraphics[width=0.55\linewidth,angle=-90]{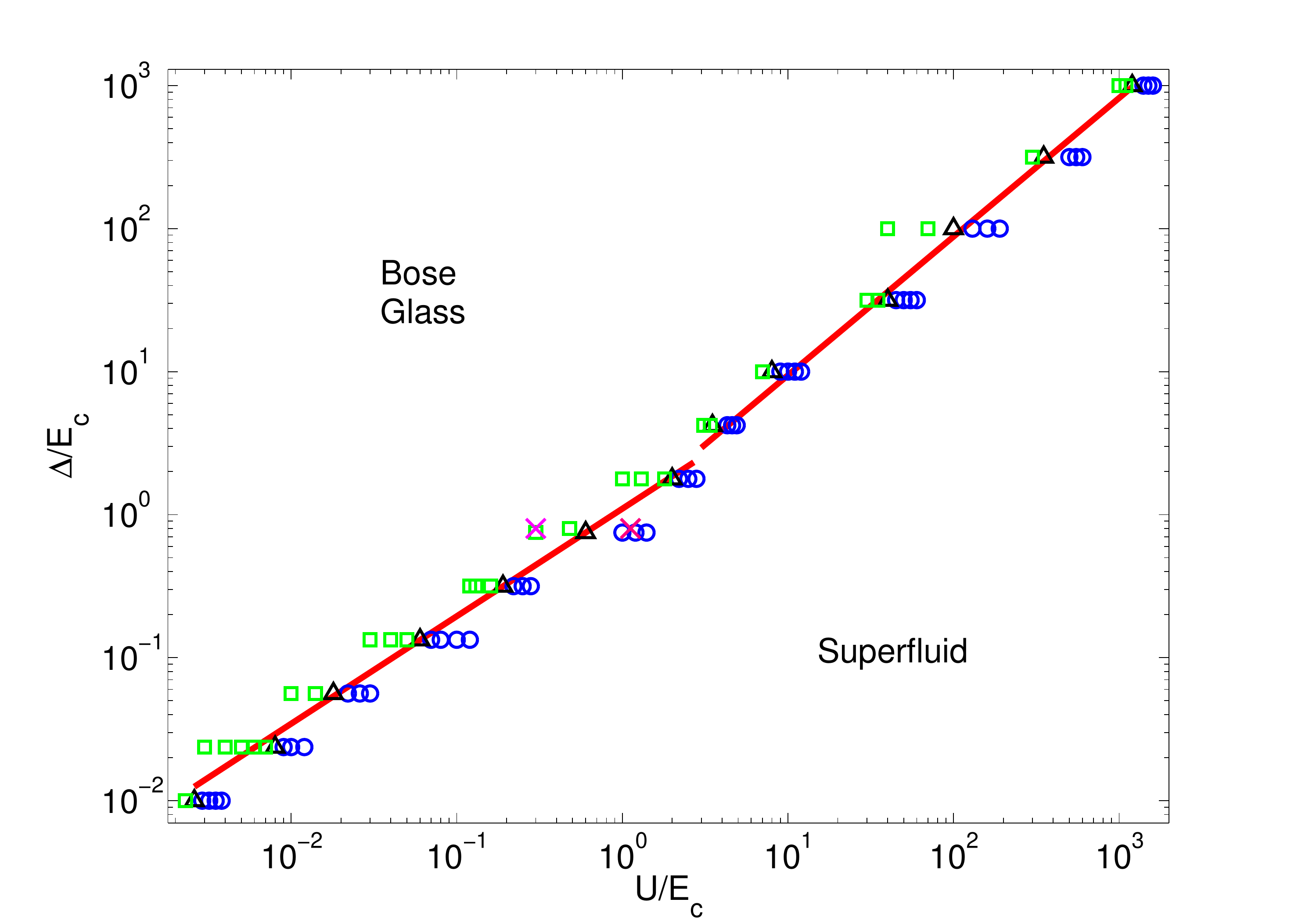}
\caption{ \label{fig:phase_diagram}
Ground-state phase diagram of weakly interacting disordered bosons in 1D. The
phases are characterized by the power-law decay (blue circles; superfluid) or
the exponential decay (green squares; Bose glass) of the averaged one-body
density matrix $\daverage{g_1}(r)$. The latter was obtained through KPM
iterations with system sizes varying between $64\,\eta$ and $2048\,\eta$. The black
triangles lie on the estimated phase boundary. Linear fits to these data points
on the double logarithmic scale yield the slope $0.75\pm 0.04$ in the
white-noise regime $U\ll E_c$, and $0.94\pm 0.03$ in the Thomas-Fermi regime
$U\gg E_c$. The purple crosses correspond to the parameters used
in the panels of Fig.~\ref{fig:BG_SF}.
}
\end{figure}

The quantum phase diagram of the weakly interacting regime can be drawn by
varying $\Delta$ and $U$, and determining for each parameter set whether the
asymptotic part of the disorder-averaged $g_1$ behaves as a power law or an
exponential. This procedure is put on a systematic footing by fitting the
long-distance part of $\daverage{g_1}(r)$ with both power laws and exponentials,
and monitoring the fit quality. Fig.~\ref{fig:phase_diagram} shows the phase
diagram obtained with the KPM technique. The blue circles and green squares have
been identified as belonging to the superfluid and Bose glass phases,
respectively. The black triangles cannot be attributed to one phase or the other
with the accuracy of the data, and are assumed to lie on the phase boundary. The
red solid lines are fits to the black triangles in the regimes $U\lesssim E_c$
and $U\gtrsim E_c$. In the white-noise regime $U\ll E_c$ (\ie $\xi\gg\eta$), the
phase boundary is expected to scale as $\Delta/E_c\sim(U/E_c)^{3/4}$, while in
the Thomas-Fermi regime $U\gg E_c$ (\ie $\xi\ll\eta$) the critical $\Delta$ is
expected to grow linearly with $U$~\cite{fontanesi_superfluid_2009,
falco_weakly_2009, fontanesi_mean-field_2010, lugan_ultracold_2010}. The fits of
Fig.~\ref{fig:phase_diagram} are in good agreement with these scaling laws. Our
findings thus reproduce the results of Refs.~\cite{fontanesi_superfluid_2009,
fontanesi_mean-field_2010} without the need for partial or complete
diagonalization. This validates our approach.

%%%%
%%%%

%%%%%%%%%%%%%%%%%%%%%%%%%%%%%%%%%%%%%%%%%%%%%%%%%%%%%%%%%
\subsection{Condensate depletion in two dimensions}
\label{subsec:2D}
%%%%%%%%%%%%%%%%%%%%%%%%%%%%%%%%%%%%%%%%%%%%%%%%%%%%%%%%%

In the absence of disorder, weakly interacting bosons form a true condensate at
$T=0$ in~2D. In this case, the reduced one-body density matrix $g_1(\vr,\vrp)$
tends to a constant equal to the condensate fraction $\rho_0/\rho$ for $\Vert
\vr-\vrp\Vert\to\infty$. To leading order in the strength of interactions, the
condensate fraction is given by~\cite{schick_two-dimensional_1971,
posazhennikova_colloquium:_2006, mora_extension_2003, muller_condensate_2012}
%%%%
\begin{equation} \label{eq:schick}
\frac{\rho_0}{\rho}
\simeq1-\frac{g'}{8\pi}
=1-\frac{1}{4\pi\rho_0 \xi^2},
\end{equation}
%%%%
where $g'=2mg/\hbar^2$ and $g$ is the 2D coupling constant. While scattering
theory shows that this constant actually depends logarithmically on the 2D
density~$\rho$ and the 3D scattering
length~$a$~\cite{posazhennikova_colloquium:_2006}, we take it as a given
parameter of Hamiltonian~(\ref{eq:HamiltonianH}) in 2D, even for the strongly
inhomogeneous case. According to Eq.~(\ref{eq:schick}), interactions reduce the
condensate fraction. In the presence of disorder, the condensate fraction is
expected to be further reduced (but nonzero) in the superfluid regime, and
completely suppressed in the Bose glass phase, due to the (exponential) decay of
the one-body density matrix.

The KPM algorithm introduced above is expected to reduce significantly the
computational cost of a study of the superfluid-insulator transition on the
basis of the one-body density matrix. A fully-fledged analysis of this
transition nevertheless lies beyond the scope of the present paper, and is left
for future work. Here, we restrict ourselves to the superfluid regime and
consider the calculation of the disorder-induced condensate depletion as a
benchmark for the KPM technique. This depletion has been calculated analytically
in Ref.~\cite{muller_condensate_2012} for the limit of weak interactions and
weak disorder. To leading order in $\Delta/U$, the depletion
$\daverage{\delta\rho}=\daverage{\rho(r)}-\daverage{\rho_0(r)}$ reads
%%%%
\begin{equation} \label{eq:depletion_analytical}
\daverage{\delta\rho}=
\delta\rho^{(0)}\left[
1+\left(\frac{\Delta}{U}\right)^2 h\left(\frac{\eta}{\xi}\right)
\right],
\end{equation}
%%%% 
where $\delta\rho^{(0)}$ is the interaction-induced condensate depletion in a
clean system, given by Eq.~(\ref{eq:schick}). The function $h$ depends only on
the ratio of the disorder correlation length~$\eta$ and the healing
length~$\xi$. Note that $h$ differs from the similar function introduced in
Ref.~\cite{muller_condensate_2012} by a trivial factor $\sqrt{2}$ in the
argument due to a different definition of $\xi$.

\begin{figure}
\centering
\includegraphics[width=0.34\linewidth,angle=-90]{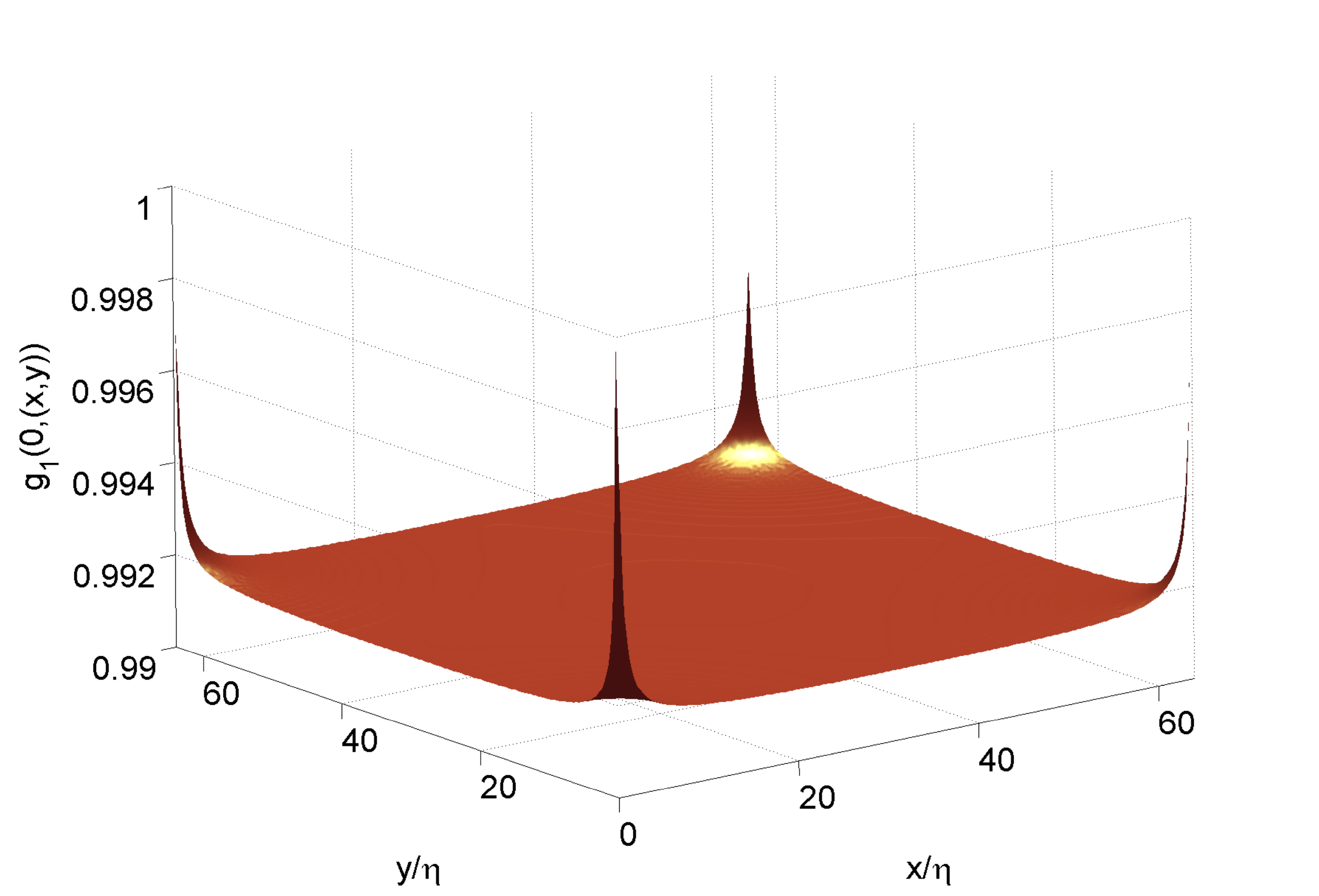}
\includegraphics[width=0.34\linewidth,angle=-90]{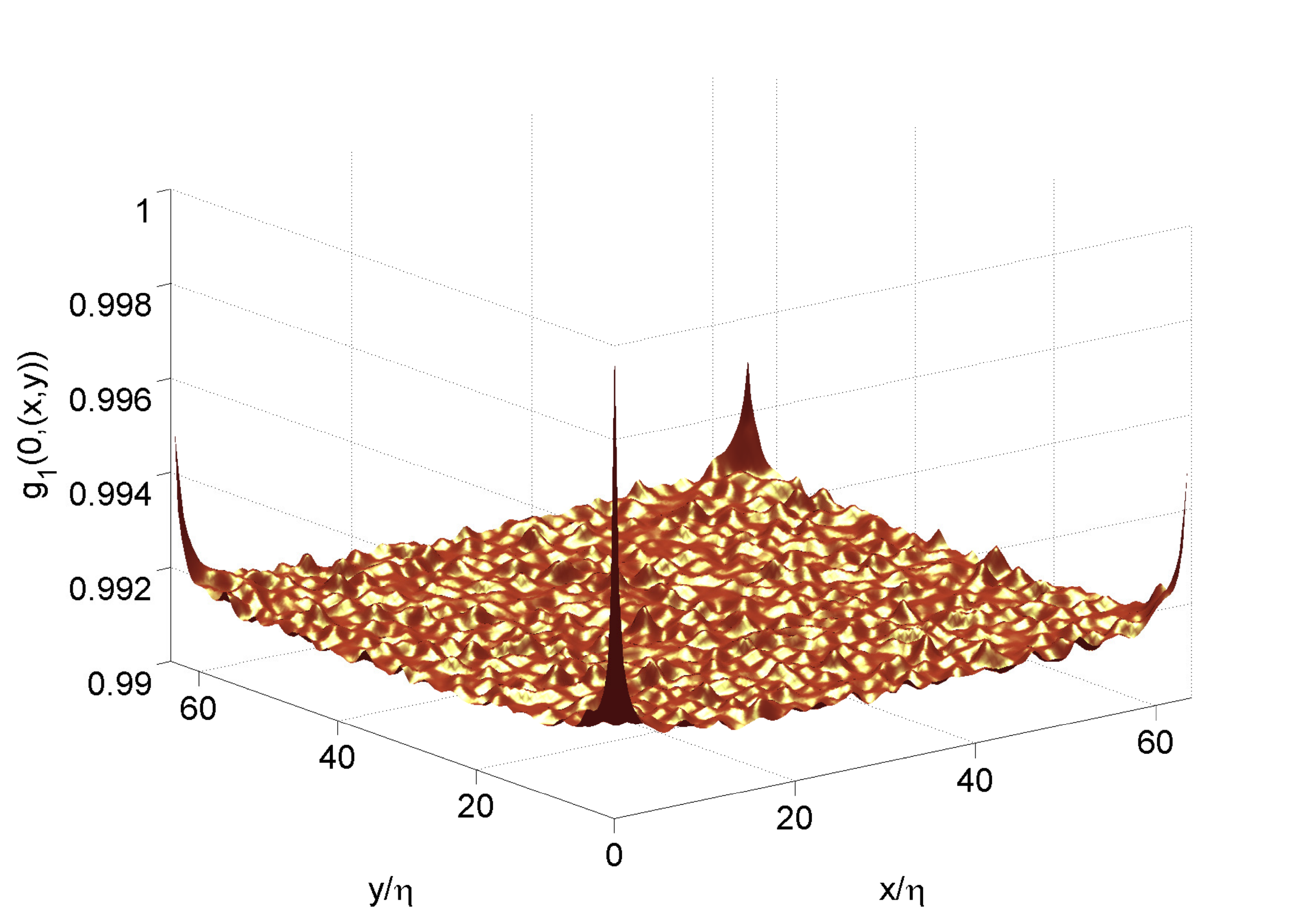}
\caption{\label{fig:2D_plateau}
Reduced one-body density matrix $g_1(\boldsymbol{0},\vr)$ for a 2D system of
size $64\eta\times 64\eta$, at $T=0$, in the absence of disorder (left panel)
and for a single disorder configuration with amplitude $\Delta=4\, E_c$ (right
panel). In both cases the interaction strength is $U=32\, E_c$ and the density
is $N_0/L^2=160\,\eta^{-2}$, which amounts to a reduced coupling constant
$g'=0.2$ [see Eq.~(\ref{eq:schick})].
}
\end{figure}

\begin{figure}
\centering
\includegraphics[width=0.55\linewidth,angle=-90]{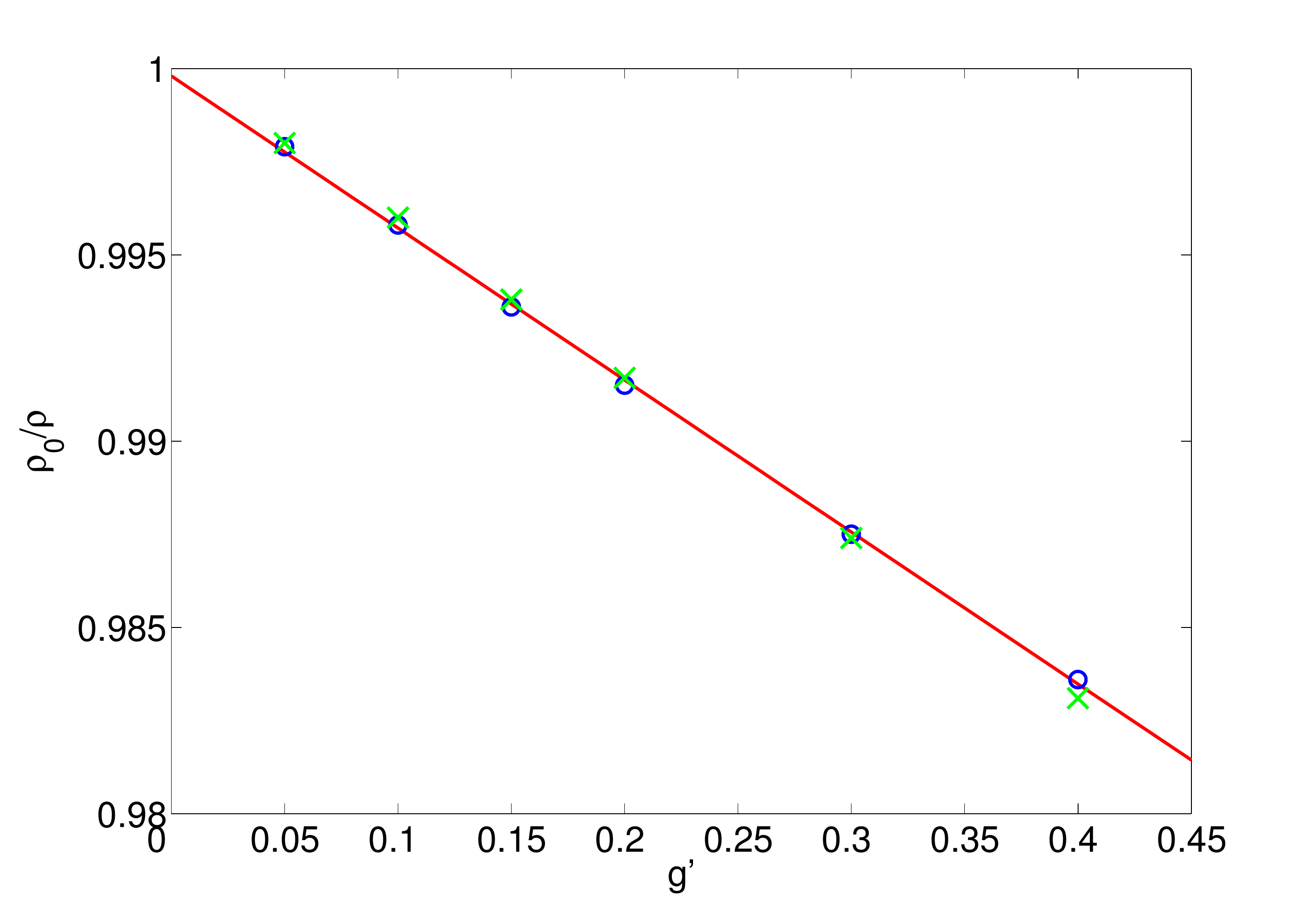}
\caption{\label{fig:schick}
Condensate fraction $\rho_0/\rho$ versus interaction strength for a 2D system.
The interaction strength is parametrized by the reduced coupling constant
$g'=2mg/\hbar^2$, where $g$ is the 2D coupling constant. The data points are the
values extracted from numerical calculations of the one-body density matrix for
particle densities $N_0/L^2=160\,\eta^{-2}$ (blue circles; system size
$64\eta\times64\eta$) and $N_0/L^2=10\,\eta^{-2}$ (green crosses; system size
$256\eta\times256\eta$), and variable interactions energies $U=gN_0/L^2$. The
solid red line is a fit to the data with $N_0/L^2=160\,\eta^{-2}$. The fitted
slope agrees with the factor $1/(8\pi)$ predicted by Eq.~(\ref{eq:schick})
within $2.5\%$. The other set of data is not fitted for the sake of clarity. The
data for $N_0/L^2=160\,\eta^{-2}$ and $g'=0.2$ corresponds the plateau in the
first panel of Fig.~\ref{fig:2D_plateau}.
}
\end{figure}

The left panel of Fig.~\ref{fig:2D_plateau} shows the result of a KPM
calculation of $g_1(\boldsymbol{0},\vr)$ in a clean 2D system. Starting from the
origin, the one-body density matrix drops and reaches a plateau beyond a few
healing lengths. The regrowth of $g_1$ on the system edges is due to the
periodic boundary conditions, as in the 1D case. The condensate fraction is
extracted from the value assumed at the center of the system. With this
procedure, we studied the dependence of the condensate fraction on the
interaction strength. The numerical results are plotted in
Fig.~\ref{fig:schick}, and stand in perfect agreement with
Eq.~(\ref{eq:schick}).

%%%%
%%%%

%%%%
%%%%

\begin{figure}
\centering
\includegraphics[width=0.55\linewidth,angle=-90]{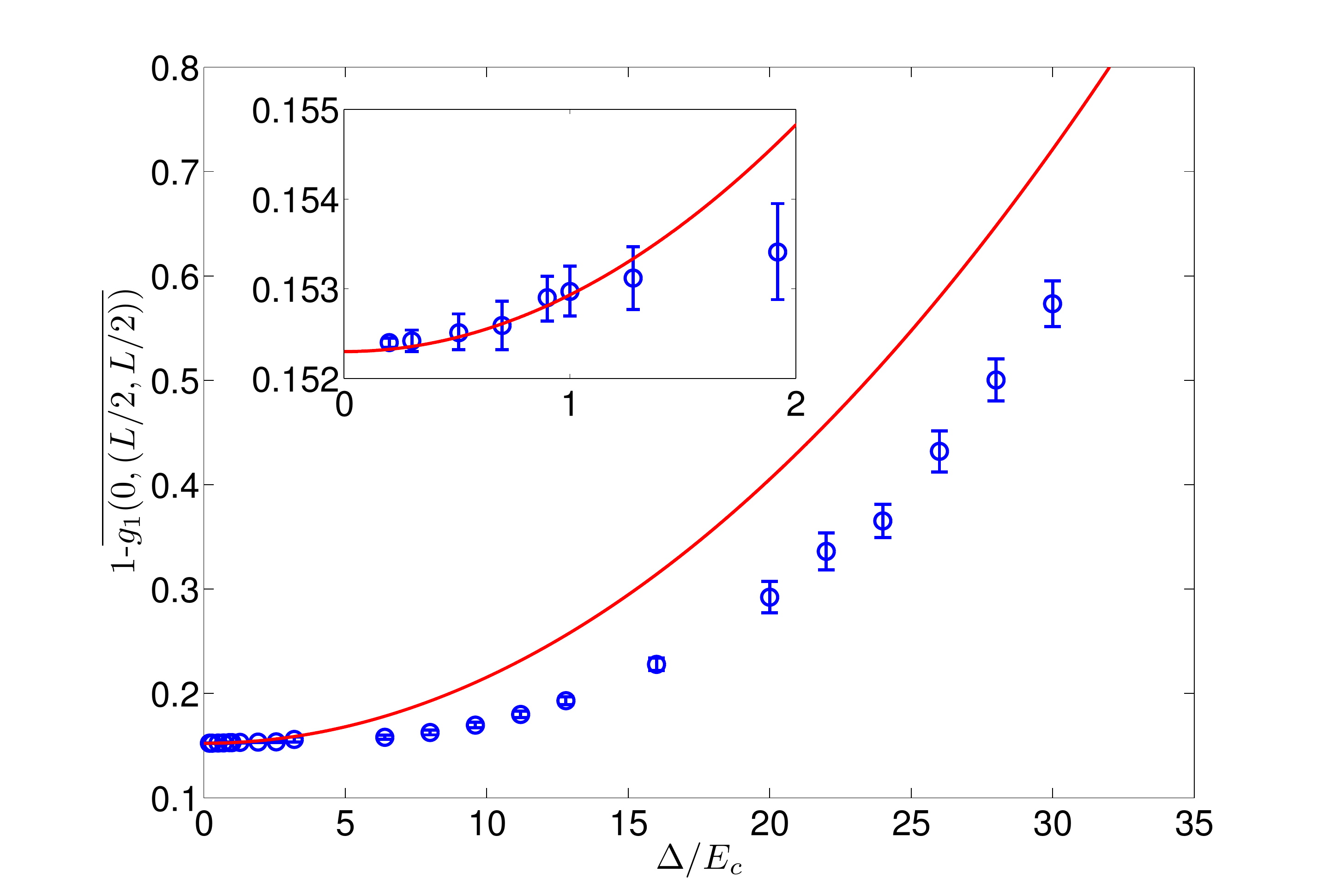}
\caption{\label{fig:depletion}
Fractional condensate depletion versus disorder strength in a system of size
$64\eta\times64\eta$, for an interaction energy $U=6.4E_c$. The numerical data
in blue represent the statistics of the reduced one-body matrix, evaluated by
KPM at a distance $\sqrt{2}L/2$ (see text) for 100 to 200 disorder
configurations. The open circles correspond to average values, and the error
bars indicate the root-mean-square fluctuations around those averages. The red
solid line represents the weak-disorder prediction
(\ref{eq:depletion_analytical}) for the depletion, normalized by the average
density.
}
\end{figure}

The disordered case was examined with a similar procedure. The right panel of
Fig.~\ref{fig:2D_plateau} displays the values of $g_1(\boldsymbol{0},\vr)$
obtained for a single disorder configuration with $\Delta/U=0.125$. For this
value the effect of disorder is weak, and the one-body density matrix still
exhibits a plateau, at roughly the same level than the clean case shown in the
left panel. As a definition of the averaged condensate fraction in the
disordered case, we use the asymptotic value of the disorder-averaged
\emph{reduced} one-body density matrix $\daverage{g_1(\vr,\vrp)}$ at large
distances. This definition follows the Penrose-Onsager criterion for
Bose-Einstein condensation, based on off-diagonal long-range
order~\cite{penrose_bose-einstein_1956}, and agrees with the definition of the
condensate fraction in the clean case. In the presence of disorder, this
definition yields a condensate fraction equal to one at the mean-field level
(where the Bose gas is described solely by the Gross-Pitaevskii equation), and
properly takes into account the role of condensate
deformation~\cite{muller_condensate_2012}. Because of this deformation of the
condensate in the presence of an inhomogeneous potential, the condensate
depletion cannot simply be associated to the fraction of atoms with momenta
$\boldsymbol{k}\neq\boldsymbol{0}$. More generally, in the presence of disorder,
the superfluid component should be characterized by spatial inhomogeneity.
Hence, the superfluid fraction is not expected to be related in a
straightforward way to the fraction of atoms with
$\boldsymbol{k}=\boldsymbol{0}$. Even if the two quantities might vanish
simultaneously at the superfluid-insulator phase boundary, the precise relation
between the condensate fraction, the superfluid fraction and the fraction of
atoms with vanishing momentum in the presence of disorder remains an open issue.
With the above definition, the statistical average and fluctuations of the
condensate fraction are evaluated by calculating $g_1(\boldsymbol{0},\vr)$,
with~$\vr$ at the system center, for several disorder configurations.
Figure~\ref{fig:depletion} shows the average and the standard deviation of the
fractional depletion $1-\lim_{r\to\infty}g_1(\boldsymbol{0},\vr)$ (\ie the
complement of the condensate fraction) as a function of the disorder strength.
For weak disorder ($\Delta\ll U$), we indeed find a good agreement with the
theoretical prediction~(\ref{eq:depletion_analytical}), as shown in the inset.
For $\Delta\gtrsim 1.3\, E_c$ (\ie $\Delta\gtrsim 0.2\, U$), however, the
numerical averages clearly lie below the theoretical curve. While differences
due to the averages used in Eq.~(\ref{eq:depletion_analytical}) and in
$\daverage{g_1}$ are not excluded, this discrepancy is likely to be due to the
breakdown of leading-order perturbation theory in the disorder amplitude. The
results of Fig.~\ref{fig:depletion} thus suggest that higher orders in the
weak-disorder expansion may reduce the condensate depletion.

%%%%%%%%%%%%%%%%%%%%%%%%%%%%%%%%%%%%%%%%%%%%%%%%%%%%%%%%%
\section{Summary and outlook}
\label{sec:conclusion}
%%%%%%%%%%%%%%%%%%%%%%%%%%%%%%%%%%%%%%%%%%%%%%%%%%%%%%%%%

We have developed an iterative scheme, based on the kernel polynomial method,
for the efficient computation of the one-body density matrix of
weakly-interacting Bose gases in the framework of Bogoliubov theory. Such a
scheme is relevant for regimes of strong disorder, which cannot be tackled
analytically. The scheme was applied to the case of disordered bosons at $T=0$
in one and two dimensions. In the one-dimensional case, we characterized the
superfluid-insulator phase transition on the basis of the long-range behavior of
the one-body density matrix, and successfully reproduced the results of
Refs.~\cite{fontanesi_superfluid_2009, fontanesi_mean-field_2010} with a low
computational overhead. In the two-dimensional geometry, we analyzed the quantum
depletion induced by interaction and disorder in the superfluid regime, and
found a good agreement with results available for the weakly disordered regime.
These case studies validate our approach and suggest that it may be used to
study the coherence properties of weakly interacting Bose systems for system
sizes that remain hardly tractable with other numerical techniques. This feature
is particularly interesting for investigations into the superfluid-insulator
transition in higher dimensions. As outlined here, our approach is also easily
extended to regimes of low but nonzero temperatures, which are relevant to
ongoing experiments with ultracold atomic gases.

%%%%%%%%%%%%%%%%%%%%%%%%%%%%%%%%%%%%%%%%%%%%%%%%%%%%%%%%%
\section*{Acknowledgments}
\label{sec:acknowledgments}
\addcontentsline{toc}{section}{Acknowledgments}
%%%%%%%%%%%%%%%%%%%%%%%%%%%%%%%%%%%%%%%%%%%%%%%%%%%%%%%%%

We would like to thank Luca Fontanesi and Michiel Wouters for fruitful discussions. This work was supported by the Swiss National Science Foundation through project no.~200020\_132407.

%%%%%%%%%%%%%%%%%%%%%%%%%%%%%%%%%%%%%%%%%%%%%%%%%%%%%%%%%
%\section*{References}
%\label{sec:references}
\addcontentsline{toc}{section}{References}
%%%%%%%%%%%%%%%%%%%%%%%%%%%%%%%%%%%%%%%%%%%%%%%%%%%%%%%%%

\bibliographystyle{iopart-num_PL}
\bibliography{Saliba_bib_20130515_arXiv-v2}

\end{document}